\documentclass[pra,aps,10pt,a4paper,showpacs,noshowkeys,notitlepage,onecolumn]{revtex4-1}

\usepackage{amsmath}
\usepackage{times}
\usepackage{amssymb}
\usepackage[dvipdfm]{graphicx}
\usepackage[unicode=true, pdfusetitle,
 bookmarks=true,bookmarksnumbered=false,bookmarksopen=false,
 breaklinks=false,pdfborder={0 0 1},backref=false,colorlinks=true,dvipdfm]
 {hyperref}

\begin{document}

\title{Controlled teleportation via photonic Faraday rotations in low-Q cavities}
\author{W. P. Bastos, W. B. Cardoso\footnote{Email address: wesleybcardoso@gmail.com}, A. T. Avelar, N. G. de Almeida and B. Baseia}
\affiliation{Instituto de F\'isica, Universidade Federal de Goi\'as, 74.001-970, Goi\^ania - GO, Brazil}

\begin{abstract}
This paper presents feasible experimental schemes to realize controlled teleportation protocols via photonic Faraday rotations in low-Q cavities. The schemes deal with controlled teleportation of superposition states and two-particle entanglement of atomic states. The information is encoded in three-level atoms in a lambda configuration trapped inside coupled cavities by optical fibers. Also, we estimate the success probability and the current feasibility of the schemes.
\end{abstract}

\maketitle

\section{Introduction}

The quantum teleportation was firstly suggested by Bennett \textit{et al.} \cite{Bennett93} and its experimental realizations reported from 1997 onwards \cite{BouwmeesterNAT97,Boschi98}. On the other hand, teleportation remains a challenge in some contexts, such as in trapped wave fields inside high-Q microwave cavities \cite{DavidovichPRA94,CiracPRA94,AlmeidaPRA00,PiresPRA04,DeSouza09PhysA}, in atomic state via cavity decay \cite{BosePRL99,YuPRA04,ChoPRA04,ChimczakPRA07,ChenJPB10}, in a the single-mode thermal state of light fields \cite{ZhangJOB05}, in trapped field states inside a single bimodal cavity \cite{PiresPRA05,QueirosPRA07}, in schemes without the Bell-state measurement \cite{ZhengPRA04,CardosoPRA05,CardosoIJTP08,dSouzaPA09,dSouzaOC11,CardosoJPB07}, in the angular spectrum of a single-photon field \cite{WalbornPRA07}, among others.

Since the pioneering work by Bennett \textit{et al.} \cite{Bennett93}, several schemes for teleportation that differ from this original protocol have appeared in the literature. As examples, in the experiment of Ref. \cite{Boschi98}, Boschi \textit{et al.} explored both the polarization and the state of the photon via two distinct paths to demonstrate teleportation; in Ref. \cite{Popescu94}, Popescu substituted the nonlocal channel by the mixed Werner states. Entanglements of mixed states as nonlocal channels were further considered in Ref. \cite{nonlocalchannelmixed}. In Ref. \cite{Vaidman94}, Vaidman proposed a \textquotedblleft cross measurement\textquotedblright\ method to achieve a two-way teleportation using a spin state and a system with continuous variable. Related to this topic, Moussa showed in \cite{Moussa97} how to implement teleportation with \textit{identity interchange} of quantum states, a kind of \textquotedblleft cross measurement\textquotedblright\ in the context of QED cavity. In Ref. \cite{deAlmeida98}, de Almeida \textit{et al.} used Greenberger-Horne-Zeilinger (GHZ) states as the nonlocal channel instead of the standard Einstein-Podolsky-Rosen (EPR) states, and in Ref. \cite{Agrawal06} Agrawal and Pati showed how to achieve perfect teleportation and superdense code using W states as the nonlocal channel. The generation of W states and clusters states in QED cavity are discussed in \cite{BecerraJPB08}. In Ref. \cite{ZhengPRA04}, Zheng refers to the approximated teleportation without Bell states measurements of the superposition of zero- and one-photon states from one high-$Q$ cavity to another, with fidelity near $99\%$ \cite{comment}. In Ref. \cite{Karlsson98}, Karlsson and Bourennane, also using a GHZ channel, showed how to accomplish teleportation controlled by a third party. Controlled teleportations involving many agents were considered in Ref. \cite{Controlled}. Since then, controlled operations have found important applications, as for example, quantum secret sharing, introduced by Hillery \textit{et al.} in Ref. \cite{Hillery99}, and experimentally reported in \cite {ExperimentalQS}. 

Quantum secret sharing was considered further by a number of authors \cite {SeveralQS}, including a version of controlled quantum secret \cite {ControlledQS}. Another important application of controlled operation by a third part is given by partial optimal teleportation, introduced in Ref. \cite {Filip04} by Filip, considered further in Ref. \cite{PartialOptimal}, and experimentally reported in Ref. \cite{PartialoptimalExp}. More recently, our group introduced the concept of controlled partial teleportation (CPT) and presented a feasible scheme for its implementation in the trapped ions domain \cite{CardosoPS09}.

In this paper, motivated by growing applications found by controlled operations \cite{Karlsson98,Controlled,
Hillery99,ExperimentalQS,SeveralQS,ControlledQS,Filip04,PartialOptimal,PartialoptimalExp}, we propose several schemes to experimentally realize controlled teleportation (CT) as well as CPT in the context of lossy optical cavities connected by optical fibers, taking advantage of the so called photonic Faraday rotations \cite{AnPRA09}. The main idea is to make use of the Faraday rotation produced by single-photon-pulse input and output process with regard to low-Q cavities \cite{JulsgaardNAT01}. In view of our applications, we revisited the input-output relation for a cavity coherently interacting with a trapped three-level atom, recently considered in Ref. \cite{AnPRA09,ChenJPB10}. We consider a three-level atom interacting with two degenerate cavity-modes of a low-Q cavity pumped by photonic emission of a single photon source via optical fibers. Fig. \ref{F1} shows the atomic levels of each atom trapped inside one of the cavities. Each transition is governed by the Jaynes-Cummings model.

The paper is organized as follows: in Sec. II we present the theoretical model, in Sec. III we present several schemes for realizing CPT and CP, and in Sec. IV we present our conclusions.

\section{Theoretical model}

\begin{figure}[t]
\centering
\includegraphics[width=2.8cm]{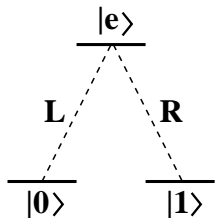}
\caption{Atomic configuration of the three-level atom trapped in each one of
the low-Q cavities. States $|0\rangle $ and $|1\rangle $ couple with a left
(L) and rigth (R) polarized photon, respectively.}
\label{F1}
\end{figure}

The Hamiltonian that describes the system of a three-level atom (Fig. \ref{F1}) interacting with a single mode of a low-Q cavity is given by \cite{Walls94}
\begin{equation}
H = H_0 + \hbar \lambda (a_L^{\dagger} \sigma_{L-}+a_L\sigma_{L+}) + \hbar \lambda (a_R^{\dagger} \sigma_{R-}+a_R\sigma_{R-}) + H_R,
\end{equation}
with
\begin{equation}
H_0 = \frac{\hbar \omega_0}{2}(\sigma_{Lz} + \sigma_{Rz}) + \hbar \omega_c (a_L^{\dagger}a_L+a_R^{\dagger}a_R),
\end{equation}
and
\begin{eqnarray}
H_R &=& H_{R0} + i\hbar \left[ \int_{-\infty}^{\infty}d\omega \sum_{j=L,R} \alpha(\omega)\left( b_j^{\dagger}(\omega)a_j + b_j(\omega)a_j^{\dagger}  \right) \right. \nonumber\\
&+& \left. \int_{-\infty}^{\infty}d\omega \sum_{j=L,R} \overline{\alpha}(\omega)\left( c_j^{\dagger}(\omega)\sigma_{j-} + c_j(\omega)\sigma_{j+}  \right) \right] ,
\end{eqnarray}
where $\lambda$ is the atom-field coupling constant, $a_j^{\dagger}$ ($a_j$) is the creation (annihilation) operator of the field-mode into the cavity with $j=L,R$, $\omega_0$ ($\omega_c$) is the atomic (field) frequency, and $\sigma_{L-}$ and $\sigma_{L+}$ ($\sigma_{R-}$ and $\sigma_{R+}$) are the lowering and raising operators of the transition L (R), respectively. The L and R transitions are shown in Fig. \ref{F1}. $H_{R0}$ is the Hamiltonian of the free reservoirs, such that, the field and atomic reservoirs are given by $H_{Rc}=\hbar\int_{-\infty}^{\infty}d\omega \omega b_j^{\dagger}b_j$ and $H_{RA}=\hbar\int_{-\infty}^{\infty}d\omega \omega c_j^{\dagger}c_j$ (j=L,R), respectively. The reservoirs couple with field and atomic systems independently, at different values of frequency $\omega$, with coupling amplitudes $\alpha=\sqrt{\kappa/2\pi}$ and $\overline{\alpha}=\sqrt{\gamma/2\pi}$, respectively. $\kappa$ and $\gamma$ are the cavity-field and atomic damping rates, $b_j$ and $c_j$ ($b_j^{\dagger}$ and $c_j^{\dagger}$) are the annihilation (creation) operators of the reservoirs.

Next, due to the fact of the presence of a pumping field into the cavity by a optical fiber one can change, in a convenient way, to a rotating frame with respect the pumping field frequency $\omega_p$ using the following transformation:
\begin{equation}
H_{eff}=U^{\dagger}HU- \left[\hbar\omega_p (a_L^{\dagger}a_L + a_R^{\dagger}a_R ) + \frac{\hbar \omega_p}{2}(\sigma_{Lz} + \sigma_{Rz}) \right],
\end{equation}
where $U=\exp\{ -i\sum_{j=L,R}[\omega_p (a_j^{\dagger}a_j + b_j^{\dagger}b_j +c_j^{\dagger}c_j)+ \frac{ \omega_p}{2}\sigma_{jz}  ]  \}$. In this point, using the Heisenberg equations for the operators $a_j$ and $\sigma_{j-}$ (consequently for $a^{\dagger}_j$ and $\sigma_{j+}$), with j=L, R, one can get
\begin{eqnarray}
\dot{a}_j(t)=-[i(\omega _{c}-\omega _{\mathrm{p}})+\frac{\kappa }{2}%
]a_j(t)-\lambda\sigma _{j-}(t)-\sqrt{\kappa }a_{\mathrm{in},j}(t),
\label{eom}
\end{eqnarray}%
\begin{eqnarray}
\dot{\sigma}_{j-}(t)&=&-[i(\omega _{0}-\omega _{\mathrm{p}})+\frac{\gamma }{2}%
]\sigma _{j-}(t) - \lambda\sigma _{jz}(t)a_j(t)+\sqrt{\gamma }\sigma _{z}(t)b_{\mathrm{in,j%
}}(t).
\label{eom2}
\end{eqnarray}%
The relation between the input and output fields reads $a_{out,j}(t)=a_{in,j}(t)+\sqrt{\kappa}a_j$, $j=$L,R. Here we work with a reservoir at zero temperature such that $b_{in,j}\simeq0$. Now, an adiabatic approximation on the above evolution equations lead us in a single relation between the input and output field states in the form \cite{AnPRA09,ChenJPB10}
\begin{equation}
r(\omega _{\mathrm{p}})=\frac{[i(\omega _{c}-\omega _{\mathrm{p}})-\frac{%
\kappa }{2}][i(\omega _{0}-\omega _{\mathrm{p}})+\frac{\gamma }{2}]+\lambda^{2}}{%
[i(\omega _{c}-\omega _{\mathrm{p}})+\frac{\kappa }{2}][i(\omega _{0}-\omega
_{\mathrm{p}})+\frac{\gamma }{2}]+\lambda^{2}},  \label{rela}
\end{equation}%
where $r(\omega _{\mathrm{p}})\equiv a_{\mathrm{out},j}(t)/a_{\mathrm{in%
},j}(t)$ is the reflection coefficient of the atom-cavity system. On the other hand, considering the case
of $\lambda=0$ and an empty cavity we have \cite{Walls94}
\begin{equation}
r_{0}(\omega _{\mathrm{p}})=\frac{i(\omega _{c}-\omega _{\mathrm{p}})-\frac{%
\kappa }{2}}{i(\omega _{c}-\omega _{\mathrm{p}})+\frac{\kappa }{2}}.
\label{r0}
\end{equation}

According to \cite{AnPRA09,ChenJPB10} the transitions $|e\rangle
\leftrightarrow |0\rangle $ and $|e\rangle \leftrightarrow |1\rangle
$ are due to the coupling to two degenerate cavity modes
$a_{\mathrm{L}}$ and $a_{\mathrm{R}}$ with left (\textrm{L}) and
right (\textrm{R}) circular polarization, respectively. For the atom initially prepared in $|0\rangle $, the transition $|0\rangle \rightarrow |e\rangle $ will occur only if the \textrm{L} circularly polarized single-photon pulse $|\mathrm{L}\rangle $ enters the cavity. Hence Eq. (\ref{rela}) leads the input pulse to the output one as $%
|\Psi _{\mathrm{out}}\rangle _{\mathrm{L}}=r(\omega _{\mathrm{p}})|\mathrm{L}%
\rangle \approx e^{i\phi }|\mathrm{L}\rangle $ with $\phi $ the
corresponding phase shift being determined by the parameter values.  
Note that an input \textrm{R} circularly polarized
single-photon pulse $|\textrm{R}\rangle $ would only sense the empty
cavity; as a consequence the corresponding output governed by Eq. (\ref{r0}) is $|\Psi _{\mathrm{out}}\rangle _{\mathrm{R}%
}=r_{0}(\omega _{\mathrm{p}})|\mathrm{R}\rangle =e^{i\phi _{0}}|\mathrm{R}%
\rangle $ with $\phi_{0}$ a phase shift different from $\phi$. Therefore,
for an input linearly polarized photon pulse $|\Psi _{\mathrm{in}}\rangle =%
\frac{1}{\sqrt{2}}(|\mathrm{L}\rangle +|\mathrm{R}\rangle )$, the
output pulse is
\begin{equation}
|\Psi _{\mathrm{out}}\rangle _{-}=\frac{1}{\sqrt{2}}(e^{i\phi }|\mathrm{L}%
\rangle +e^{i\phi _{0}}|\mathrm{R}\rangle ).
\label{fr1}
\end{equation}%
This also implies that the polarization direction of the reflected photon rotates an angle $\Theta _{F}^{-}=(\phi_0 -\phi)/2$ with respect to that of the input one, called Faraday rotation \cite{JulsgaardNAT01}. If the atom is initially prepared in $|1\rangle $, then only the $\mathrm{R}$ circularly polarized photon could sense the atom, whereas the $\mathrm{L}$ circularly polarized photon only interacts with the empty cavity. So we have,
\begin{equation}
|\Psi _{\mathrm{out}}\rangle _{+}=\frac{1}{\sqrt{2}}(e^{i\phi _{0}}|\mathrm{L}\rangle +e^{i\phi }|\mathrm{R}\rangle ),
\label{fr2}
\end{equation}
where the Faraday rotation is $\Theta _{F}^{+}=(\phi -\phi _{0})/2$.


\section{Controlled teleportation}

In this Section we present three cases of CT. Firstly we will consider CT of superposition states. Secondly, we present CPT of entangled states and finally we present CT of entangled states.

\subsection{Superposition states}
\noindent

Here, we consider the CT of superposition states with one, two, or more controls. 

\begin{figure}[t]
\centering
\includegraphics[width=7cm]{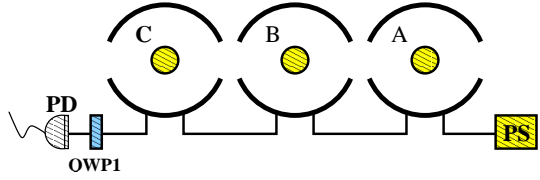}
\caption{Scheme for teleporting the superposed atomic state $C$ using one
control $B$. $PS$ stands for the photon source, $A$, $B$, and $C$ represent the
atoms trapped inside the corresponding cavities. $QWP1$ and $PD$ are quarter
wave plate and polarization photodetector, respectively.}
\label{F2}
\end{figure}

\textit{i}) \textit{One control}: in this scheme we deal with three atoms each one trapped inside three low-Q cavities ($A$, $B$, and $C$), respectively. The atoms are previously prepared in specific states, such that, $|\psi_{A}\rangle=\frac{1}{\sqrt{2}}(|0\rangle_{A}+|1\rangle_{A})$, $|\psi_{B}\rangle=\frac{1}{\sqrt{2}}(|0\rangle_{B}+|1\rangle_{B})$, $|\psi_{C}\rangle=\alpha|0\rangle_{C}+\beta|1\rangle_{C}$, where $\alpha$ and $\beta$ are unknown coefficients that obey $|\alpha|^2+|\beta|^2=1$, and the subindex represents the atom trapped in the cavity. Also, we consider a linearly polarized photon in the state $|\psi_{p}\rangle=\frac{1}{\sqrt{2}}(|L\rangle+|R\rangle)$, where $|L\rangle$ ($|R\rangle$) represents the state with left (right) direction of polarization. The experimental scheme is displayed in Fig. \ref{F2}.

The nonlocal channel is created after the interaction of the photon with the atom trapped inside the cavity A. Due to the low quality of the cavity the photon is lost, escaping through an optical fiber directed to the cavity B. This interaction causes a Faraday rotation in the photonic state (see Eqs. (\ref{fr1}) and (\ref{fr2})) leading the entire atom-photon state to
\begin{equation}
|\phi_{1}\rangle=\frac{1}{2}(e^{i\phi}|L0\rangle_{A}+e^{i\phi_{0}}|L1\rangle_{A}+e^{i\phi_{0}}|R0\rangle_{A}+e^{i\phi}|R1\rangle_{A}),
\end{equation}
where the phases $\phi$ and $\phi_0$ are obtained by the reflection coefficients in Eqs. (\ref{rela}) and (\ref{r0}). 

After the photon has interacted with the atom trapped inside the cavity B and considering the adjustments $\omega_{p}=\omega_{c}-\kappa/2$, $g=\kappa/2$ and $\omega_{0}=\omega_{c}$ (consisting in $\phi=\pi$ and $\phi_{0}=\pi/2$), we have
\begin{align}
|\phi_{2}\rangle=&\frac{1}{2\sqrt{2}}\Big(|0\rangle_{B}\big[|L\rangle|0\rangle_{A}-i|L\rangle|1\rangle_{A}-|R\rangle|0\rangle_{A}-i|R\rangle|1\rangle_{A}   \big]+\notag\\
&|1\rangle_{B}\big[-i|L\rangle|0\rangle_{A}-|L\rangle|1\rangle_{A}-i|R\rangle|0\rangle_{A}+|R\rangle|1\rangle_{A}   \big]\Big).
\end{align}
\label{p1}

In the following, the photon interacts with the atom trapped inside the cavity C leading the whole state as
\begin{align}
|\phi_{3}\rangle=&\frac{1}{2\sqrt{2}}\Big(|0\rangle_{B}\big[-\alpha(|L\rangle+i|R\rangle)|0\rangle_{A}|0\rangle_{C}+i\alpha(|L\rangle-i|R\rangle)|1\rangle_{A}|0\rangle_{C}\notag\\
& +i\beta(|L\rangle-i|R\rangle)|0\rangle_{A}|1\rangle_{C}+\beta(|L\rangle+i|R\rangle)|1\rangle_{A}|1\rangle_{C}     \big]\notag\\
&+|1\rangle_{B}\big[i\alpha(|L\rangle-i|R\rangle)|0\rangle_{A}|0\rangle_{C}+\alpha(|L\rangle+i|R\rangle)|1\rangle_{A}|0\rangle_{C}\notag\\
& +\beta(|L\rangle+i|R\rangle)|0\rangle_{A}|1\rangle_{C}-i\beta(|L\rangle-i|R\rangle)|1\rangle_{A}|1\rangle_{C}     \big]\Big)
\label{eqa1}
\end{align}

Next, the photon goes through a quarter wave plate (QWP1 in Fig. 2), suffering a rotation in the polarization state (Hadamard operation) such that
\begin{subequations}
\begin{eqnarray}
(|L\rangle+i|R\rangle)\sqrt{2}\rightarrow|L\rangle,\\
(|L\rangle-i|R\rangle)\sqrt{2}\rightarrow|R\rangle.
\end{eqnarray}
\label{s1}
\end{subequations}
Besides, including a Hadamard operation in the state of the atom C, the system evolves to
\begin{align}
|\phi_{4}\rangle=&\frac{1}{2\sqrt{2}}\Big(|0\rangle_{B}\Big[-|L0\rangle_{C}(\alpha|0\rangle_{A}-\beta|1\rangle_{A}) -|L1\rangle_{C}(\alpha|0\rangle_{A}+\beta|1\rangle_{A}) \notag\\
&+i|R0\rangle_{C}(\alpha|1\rangle_{A}+\beta|0\rangle_{A})+i|R1\rangle_{C}(\alpha|1\rangle_{A}-\beta|0\rangle_{A}) \Big]+\notag\\
&|1\rangle_{B}\Big[|L0\rangle_{C}(\alpha|1\rangle_{A}+\beta|0\rangle_{A})+|L1\rangle_{C}(\alpha|1\rangle_{A}-\beta|0\rangle_{A})+\notag\\
&i|R0\rangle_{C}(\alpha|0\rangle_{A}-\beta|1\rangle_{A})+i|R1\rangle_{C}(\alpha|0\rangle_{A}+\beta|1\rangle_{A})\Big] \Big)
\end{align}
Then, by measuring the photon polarization state plus the state of the atoms trapped inside the cavities B and C it is possible the reconstruction of the teleported state via appropriated atomic rotations, as summarized in Table \ref{T1}. Note that the teleportation is concluded if and only if the result of the state of the atom trapped inside the cavity B is known in terminal A. So, this atom is here treated as an agent that control the teleportation scheme.

\begin{table}
\begin{center}
\begin{tabular}{|| c | c | c | c ||}
\hline MAPS & CS & TR & AO \\ 
\hline  $|L0\rangle_{FC}$  & $|0\rangle_{B}$ & $\alpha|0\rangle_{A}-\beta|1\rangle_{A}$ & $\sigma_z$ \\ 
\hline  $|L1\rangle_{FC}$  & $|0\rangle_{B}$ & $\alpha|0\rangle_{A}+\beta|1\rangle_{A}$ & $\mathbb{I}$ \\  
\hline  $|R0\rangle_{FC}$  & $|0\rangle_{B}$ & $\alpha|1\rangle_{A}+\beta|0\rangle_{A}$ & $\sigma_{x}$ \\  
\hline  $|R1\rangle_{FC}$  & $|0\rangle_{B}$ & $\alpha|1\rangle_{A}-\beta|0\rangle_{A}$ & $\sigma_z\sigma_x$ \\ 
\hline  $|L0\rangle_{FC}$  & $|1\rangle_{B}$ & $\alpha|1\rangle_{A}+\beta|0\rangle_{A}$ & $\sigma_{x}$ \\ 
\hline  $|L1\rangle_{FC}$  & $|1\rangle_{B}$ & $\alpha|1\rangle_{A}-\beta|0\rangle_{A}$ & $\sigma_z\sigma_x$ \\  
\hline  $|R0\rangle_{FC}$  & $|1\rangle_{B}$ & $\alpha|0\rangle_{A}-\beta|1\rangle_{A}$ & $\sigma_z$ \\  
\hline  $|R1\rangle_{FC}$  & $|1\rangle_{B}$ & $\alpha|0\rangle_{A}+\beta|1\rangle_{A}$ & $\mathbb{I}$ \\  
\hline 
\end{tabular}
\end{center}
\caption{Possible results and rotations for completing the controlled teleportation procedure for the case \textit{i}). The first column shows the possible results of measurements on the atom $C$ and photon $F$ states. Second and third columns do the same for the control state ($CS$) and for the teleported ($TR$). Fourth column shows the corresponding Pauli matrices representing unitary operations upon the atomic state ($AO$) required to complete the teleportation process.}
\label{T1}
\end{table}

\begin{figure}[t]
\centering
\includegraphics[width=8.5cm]{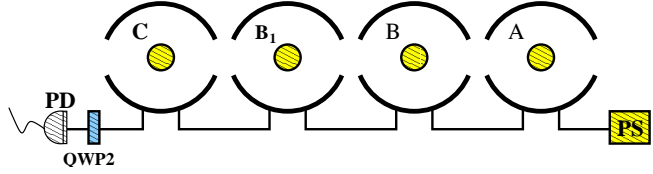}
\caption{Schematic representation to teleport a superposed atomic state
with two controls. $PS$ is the photon source, $A$, $B$, $B_{1}$, and $C$
represents the atoms trapped in the cavities, $QWP2$ is a quarter wave
plate, and $PD$ is the photodetector of polarization.}
\label{F3}
\end{figure}

\textit{ii) Two controls}: This scheme is similar to those shown above. Here, an atom, which will work as a new control, is added to the new cavity ($B_1$). One atom is previously prepared in the state $|\psi_{B_1}\rangle=\frac{1}{\sqrt{2}}(|0\rangle_{B_1}+|1\rangle_{B_1})$ while the others are as before. The experimental setup is displayed in Fig. \ref{F3}. After the photon to interact with the atoms trapped inside cavities $A$ and $B$, respectively, using the same adjustments considered in the previous Section, the state of the system results as that in Eq. (12). The next step consists in the interaction of the photon, outgoing the cavity $B$, with the atom trapped in the cavity $B_1$. So, after the photonic Faraday rotation the state of the system is given by
\begin{align}
|\phi_{3}\rangle=&\frac{1}{4}\Big(|00\rangle_{B B_{1}}\big(-|L0\rangle_{A} + i|L1\rangle_{A} - i|R0\rangle_{A} + |R1 \rangle_{A}\big)\notag\\
&+|01\rangle_{B B_{1}}\big(i|L0\rangle_{A} + |L1\rangle_{A} + |R0\rangle_{A} + i|R1 \rangle_{A}\big)\notag\\
&+|10\rangle_{B B_{1}}\big(i|L0\rangle_{A} + |L1\rangle_{A} + |R0\rangle_{A} + i|R1 \rangle_{A}\big)\notag\\
&-|11\rangle_{B B_{1}}\big(-|L0\rangle_{A} + i|L1\rangle_{A} - i|R0\rangle_{A} + |R1 \rangle_{A}\big)\Big).
\end{align}

Then, this photon is left to interact with the atom trapped inside the cavity $C$, in a way such that the whole state of the system takes the form
\begin{align}
|\phi_{3}^{'}\rangle=&\frac{1}{4}\Big(|00\rangle_{B B_{1}}\Big[\alpha\big(|L\rangle + |R\rangle\big)|0\rangle_{A}|0\rangle_{C} -i\alpha\big(|L\rangle - |R\rangle\big)|1\rangle_{A}|0\rangle_{C}-i\beta\big(|L\rangle - |R\rangle\big)|0\rangle_{A}|1\rangle_{C}-\beta(|L\rangle + |R\rangle)|1\rangle_{A}|1\rangle_{C}  \Big]\notag\\
&+|01\rangle_{B B_{1}}\Big[-i\alpha\big(|L\rangle - |R\rangle\big)|0\rangle_{A}|0\rangle_{C} - \alpha\big(|L\rangle + |R\rangle\big)|1\rangle_{A}|0\rangle_{C}-\beta\big(|L\rangle + |R\rangle\big)|0\rangle_{A}|1\rangle_{C}+i\beta\big(|L\rangle - |R\rangle\big)|1\rangle_{A}|1\rangle_{C}  \Big]\notag\\
&+|10\rangle_{B B_{1}}\Big[-i\alpha\big(|L\rangle - |R\rangle\big)|0\rangle_{A}|0\rangle_{C} - \alpha\big(|L\rangle + |R\rangle\big)|1\rangle_{A}|0\rangle_{C}-\beta\big(|L\rangle + |R\rangle\big)|0\rangle_{A}|1\rangle_{C}+i\beta\big(|L\rangle - |R\rangle\big)|1\rangle_{A}|1\rangle_{C}  \Big]\notag\\
&-|11\rangle_{B B_{1}}\Big[\alpha\big(|L\rangle + |R\rangle\big)|0\rangle_{A}|0\rangle_{C} -i\alpha\big(|L\rangle - |R\rangle\big)|1\rangle_{A}|0\rangle_{C}-i\beta\big(|L\rangle - |R\rangle\big)|0\rangle_{A}|1\rangle_{C}-\beta\big(|L\rangle + |R\rangle\big)|1\rangle_{A}|1\rangle_{C}  \Big]\Big). 
\end{align}

Next, the photon that leaves the cavity C crosses a quarter wave plate ($QWP2$ in Fig. \ref{F3}) and its state results in
\begin{subequations}
\begin{eqnarray}
(|L\rangle+|R\rangle)\sqrt{2}\rightarrow|L\rangle,\\
(|L\rangle-|R\rangle)\sqrt{2}\rightarrow|R\rangle.
\end{eqnarray}
\label{s2}
\end{subequations}
A Hadamard operation in the atomic state of the C, we have
\begin{align}
|\phi_{4}\rangle=&\frac{1}{4}\Big(|00\rangle_{BB_{1}}\Big[|L0\rangle_{FC}(\alpha|0\rangle_{A}-\beta|1\rangle_{A})+|L1\rangle(\alpha|0\rangle_{A}+\beta|1\rangle_{A})\notag\\
&-i|R0\rangle_{FC}(\alpha|1\rangle_{A}+\beta|0\rangle_{A})-i|R1\rangle_{FC}(\alpha|1\rangle_{A}-\beta|0\rangle_{A})  \Big]\notag\\ 
&+|01\rangle_{BB_{1}}\Big[-|L0\rangle_{FC}(\alpha|1\rangle_{A}+\beta|0\rangle_{A})-|L1\rangle_{FC}(\alpha|1\rangle_{A}-\beta|0\rangle_{A})\notag\\
&-i|R0\rangle_{FC}(\alpha|0\rangle_{A}+\beta|1\rangle_{A})-i|R1\rangle_{FC}(\alpha|0\rangle_{A}-\beta|1\rangle_{A})  \Big]\notag\\
&+|10\rangle_{BB_{1}}\Big[-|L0\rangle_{FC}(\alpha|1\rangle_{A}+\beta|0\rangle_{A})-|L1\rangle_{FC}(\alpha|1\rangle_{A}-\beta|0\rangle_{A})\notag\\
&-i|R0\rangle_{FC}(\alpha|0\rangle_{A}+\beta|1\rangle_{A})-i|R1\rangle_{FC}(\alpha|0\rangle_{A}-\beta|1\rangle_{A})  \Big]\notag\\
&+|11\rangle_{BB_{1}}\Big[-|L0\rangle_{FC}(\alpha|0\rangle_{A}-\beta|1\rangle_{A})-|L1\rangle_{FC}(\alpha|0\rangle_{A}+\beta|1\rangle_{A})\notag\\
&+i|R0\rangle_{FC}(\alpha|1\rangle_{A}+\beta|0\rangle_{A})+i|R1\rangle_{FC}(\alpha|1\rangle_{A}-\beta|0\rangle_{A})  \Big] \Big).
\end{align}
The appropriate operations influenced by the measurement on the control systems for the teleported state are summarized in the Table \ref{T2}.

\begin{table}
\begin{center}
\begin{tabular}{|| c | c | c | c |}
\hline MAPS & TR & TR & AO \\ 
\hline  $|L0\rangle_{FC}$  & $|00\rangle_{BB_{1}}$ & $\alpha|0\rangle_{A}-\beta|1\rangle_{A}$ & $\sigma_z$ \\ 
\hline  $|L1\rangle_{FC}$  & $|00\rangle_{BB_{1}}$ & $\alpha|0\rangle_{A}+\beta|1\rangle_{A}$ & $\mathbb{I}$ \\  
\hline  $|R0\rangle_{FC}$  & $|00\rangle_{BB_{1}}$ & $\alpha|1\rangle_{A}+\beta|0\rangle_{A}$ & $\sigma_{x}$ \\  
\hline  $|R1\rangle_{FC}$  & $|00\rangle_{BB_{1}}$ & $\alpha|1\rangle_{A}-\beta|0\rangle_{A}$ & $\sigma_z\sigma_x$ \\ 
\hline  $|L0\rangle_{FC}$  & $|01\rangle_{BB_{1}}$ & $\alpha|1\rangle_{A}+\beta|0\rangle_{A}$ & $\sigma_x$ \\ 
\hline  $|L1\rangle_{FC}$  & $|01\rangle_{BB_{1}}$ & $\alpha|1\rangle_{A}-\beta|0\rangle_{A}$ & $\sigma_z\sigma_x$ \\  
\hline  $|R0\rangle_{FC}$  & $|01\rangle_{BB_{1}}$ & $\alpha|0\rangle_{A}+\beta|1\rangle_{A}$ & $\mathbb{I}$ \\  
\hline  $|R1\rangle_{FC}$  & $|01\rangle_{BB_{1}}$ & $\alpha|0\rangle_{A}-\beta|1\rangle_{A}$ & $\sigma_z$ \\ 
\hline 
\end{tabular}
\begin{tabular}{| c | c | c | c ||}
\hline MAPS & TR & TR & AO \\ 
\hline  $|L0\rangle_{FC}$  & $|10\rangle_{BB_{1}}$ & $\alpha|1\rangle_{A}+\beta|0\rangle_{A}$ & $\sigma_x$ \\ 
\hline  $|L1\rangle_{FC}$  & $|10\rangle_{BB_{1}}$ & $\alpha|1\rangle_{A}-\beta|0\rangle_{A}$ & $\sigma_z\sigma_x$ \\  
\hline  $|R0\rangle_{FC}$  & $|10\rangle_{BB_{1}}$ & $\alpha|0\rangle_{A}+\beta|1\rangle_{A}$ & $\mathbb{I}$ \\  
\hline  $|R1\rangle_{FC}$  & $|10\rangle_{BB_{1}}$ & $\alpha|0\rangle_{A}-\beta|1\rangle_{A}$ & $\sigma_z$ \\
\hline  $|L0\rangle_{FC}$  & $|11\rangle_{BB_{1}}$ & $\alpha|0\rangle_{A}-\beta|1\rangle_{A}$ & $\sigma_z$ \\ 
\hline  $|L1\rangle_{FC}$  & $|11\rangle_{BB_{1}}$ & $\alpha|0\rangle_{A}+\beta|1\rangle_{A}$ & $\mathbb{I}$ \\  
\hline  $|R0\rangle_{FC}$  & $|11\rangle_{BB_{1}}$ & $\alpha|1\rangle_{A}+\beta|0\rangle_{A}$ & $\sigma_{x}$ \\  
\hline  $|R1\rangle_{FC}$  & $|11\rangle_{BB_{1}}$ & $\alpha|1\rangle_{A}-\beta|0\rangle_{A}$ & $\sigma_z\sigma_x$ \\
\hline 
\end{tabular}
\end{center}
\caption{Possible results and rotations to complete the controlled
teleportation procedure for the case \textit{ii}). The first column shows the
possible results of measurements on the atom $C$ and photon $F$ states.
Second and third columns do the same for the two control states ($CS$)
and for the teleported ($TR$). Fourth column shows the corresponding Pauli
matrices representing unitary operations on the atomic state ($AO$) required
to complete the teleportation process.} 
\label{T2}
\end{table}

\textit{iii) Generalization}: A procedure to generalize the number of controls in the scheme of controlled teleportation of superposition states can be done by the adding of atoms trapped in additional cavities and using specific quarter wave plates after the last cavity. In this way, if the number of controls is odd, a QWP1 is required with rotations given by Eqs. (\ref{s1}a) and (\ref{s1}b) is necessary; if the control number is even one request a QWP2 with the rotations given by Eqs. (\ref{s2}a) and (\ref{s2}b). So, with appropriated rotations we can recover the teleported state with the knowledge of the control results.

\subsection{Controlled partial teleportation of entangled states}
\noindent

In our scheme for CPT \cite{CardosoPS09}, an entangled state of particle $A$, given to Alice, and particle $B$, given to Bob, is to be partially teleported. Meanwhile, a quantum channel composed by an entanglement of three particles is shared by Alice ($A^{\prime }$), Ben ($B^{\prime }$), and a third part, say Chris ( $C$). If Alice performs a Bell measurement on the states of particle $A$ and $A^{\prime }$, and Chris performs a measurement on the state of particle $C$, then when both inform Bob their results, the following interesting result emerges, after the usual rotation performed by Bob: particle $B^{\prime }$ takes exactly the role of particle $A$ in the previous entanglement shared by Alice and Bob. As the entanglement between the particles $A$ and $B$ is broken and a new entanglement between
the particles $B$ and $B^{\prime }$ is created in a different place, depending on the collaboration of both Alice and Chris, this characterizes a controlled partial teleportation. 

Next, we describe some schemes for CPT with different number of controls taking into account the low Q cavities scenario combined with Faraday rotations.

\begin{figure}[t]
\centering
\includegraphics[width=7cm]{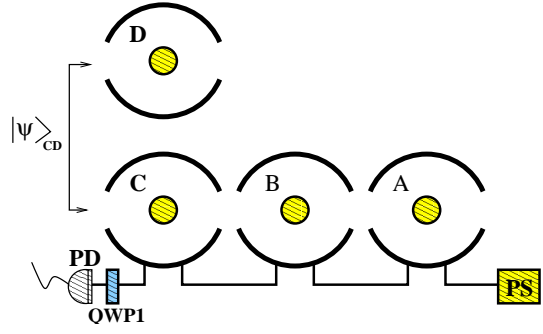}
\caption{Schematic representation for partial teleportation of entangled
atomic states with one control. $PS$ stands for the photon source, $A$, $B$, $C$,
and $D$ represents the atoms trapped inside the cavities, $QWP1$ stands for a quarter
wave plate, and $PD$ is the photodetector of polarization.}
\label{F4}
\end{figure}

\textit{i) One control}: To perform CPT, four cavities are required as displayed in Fig. \ref{F4}. Cavities A and B are previously prepared in the superposed states while the cavities C and D are previously prepared in the entangled state that we want to teleport, given by
\begin{equation}
|\psi\rangle_{CD}=\alpha|01\rangle_{CD}+\beta|10\rangle_{CD}.
\label{CD}
\end{equation}

Here we will follow the same procedure of the CT-schemes used in 3.1: the photon, previously prepared in a superposed polarization state, enters in the cavity A and interacts with the atom. Next, this photon is sent to interact with the atom trapped in the cavity B. After that, the photon interacts with the atom C in the presence of the same Faraday rotation discussed in Section A. Before the Faraday rotations, the state of the system is written as
\begin{align}
|\delta_{1}\rangle=&\frac{1}{2\sqrt{2}}\Bigg( |0\rangle_{B}\Big[-\alpha|0\rangle_{A}|01\rangle_{CD}\big(|L\rangle + i|R\rangle \big) +i\alpha|1\rangle_{A}|01\rangle_{CD}\big(|L\rangle - i|R\rangle \big)\notag\\
& +i\beta|0\rangle_{A}|10\rangle_{CD}\big(|L\rangle - i|R\rangle \big) + \beta|1\rangle_{A}|10\rangle_{CD}\big(|L\rangle + i|R\rangle \big)\Big]\notag\\
&+|1\rangle_{B}\Big[i\alpha|0\rangle_{A}|01\rangle_{CD}\big(|L\rangle - i|R\rangle \big) +\alpha|1\rangle_{A}|01\rangle_{CD}\big(|L\rangle + i|R\rangle \big)\notag\\
& +\beta|0\rangle_{A}|10\rangle_{CD}\big(|L\rangle + i|R\rangle \big)  -i\beta|1\rangle_{A}|10\rangle_{CD}\big(|L\rangle - i|R\rangle \big)\Big]\Bigg).
\end{align}

After the Faraday rotation in the atom in cavity C as well as in the photon state through the QWP1, the state in Eq. 21 goes to

\begin{align}
|\delta_{2}\rangle=&\frac{1}{2\sqrt{2}}\Bigg(|0\rangle_{B}\Big[-|L0\rangle_{C}(\alpha|01\rangle_{AD}-\beta|10\rangle_{AD})-|L1\rangle_{C}(\alpha|01\rangle_{AD}+\beta|10\rangle_{AD})\notag\\
&-i|R0\rangle_{C}(\alpha|11\rangle_{AD}+\beta|00\rangle_{AD})+i|R1\rangle_{C}(\alpha|11\rangle_{AD}-\beta|00\rangle_{AD})  \Big]+\notag\\
&|1\rangle_{B}\Big[|L0\rangle_{C}(\alpha|11\rangle_{AD}+\beta|00\rangle_{AD})+|L1\rangle_{C}(\alpha|11\rangle_{AD}-\beta|00\rangle_{AD})\notag\\
&+i|R0\rangle_{C}(\alpha|01\rangle_{AD}-\beta|10\rangle_{AD})+i|R1\rangle_{C}(\alpha|01\rangle_{AD}+\beta|10\rangle_{AD})  \Big] \Bigg),
\end{align}

Then, the controlled teleportation is concluded after the operations described in the Table \ref{T3}. Note that the knowledge of the atom state controlling the teleportation is essential to complete the process.

\begin{table}
\begin{center}
\begin{tabular}{|| c | c | c | c |}
\hline MAPS & FP & ESR & AO \\ 
\hline  $|L0\rangle_{FC}$  &  $|0\rangle_{B}$ & $\alpha|01\rangle_{AD}-\beta|10\rangle_{AD}$ & $\sigma_z\otimes\mathbb{I}$ \\
 
\hline  $|L1\rangle_{FC}$  & $|0\rangle_{B}$ & $\alpha|01\rangle_{AD}+\beta|10\rangle_{AD}$ & $\mathbb{I}\otimes\mathbb{I}$ \\

\hline  $|R0\rangle_{FC}$  & $|0\rangle_{B}$ & $\alpha|11\rangle_{AD}+\beta|00\rangle_{AD}$ & $\sigma_x\otimes\mathbb{I}$ \\

\hline  $|R1\rangle_{FC}$  & $|0\rangle_{B}$ & $\alpha|11\rangle_{AD}-\beta|00\rangle_{AD}$ & $\sigma_z\sigma_x\otimes\mathbb{I}$ \\ 

\hline  $|L0\rangle_{FC}$  & $|1\rangle_{B}$ & $\alpha|11\rangle_{AD}+\beta|00\rangle_{AD}$ & $\sigma_x\otimes\mathbb{I}$ \\
 
\hline  $|L1\rangle_{FC}$  & $|1\rangle_{B}$ & $\alpha|11\rangle_{AD}-\beta|00\rangle_{AD}$ & $\sigma_z\sigma_x\otimes\mathbb{I}$ \\

\hline  $|R0\rangle_{FC}$  & $|1\rangle_{B}$ & $\alpha|01\rangle_{AD}-\beta|10\rangle_{AD}$ & $\sigma_z\otimes\mathbb{I}$ \\

\hline  $|R1\rangle_{FC}$  & $|1\rangle_{B}$ & $\alpha|01\rangle_{AD}+\beta|10\rangle_{AD}$ & $\mathbb{I}\otimes\mathbb{I}$ \\ 
\hline 
\end{tabular}
\end{center}
\caption{Possible results and rotations to complete the controlled
partial teleportation procedure for the case \textit{i}). The first column
shows the possible results of measurements on the atom $C$ and photon $F$
states. Second and third columns do the same for the control state ($CS$%
) and for the teleported state ($TR$). Fourth column shows the corresponding
Pauli matrices representing unitary operations upon the atomic state ($AO$)
required to complete the teleportation process.}
\label{T3}
\end{table}

\begin{figure}[t]
\centering
\includegraphics[width=7cm]{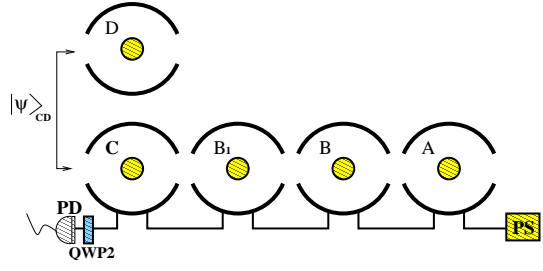}
\caption{Schematic representation for controlled partial teleportation of
entangled atomic states with two controls. $PS$ stands for photon source, $A$, $B
$, $B_{1}$, $C$, and $D$ represent the atoms trapped inside the cavities, $QWP2$
is for quarter wave plate, and $PD$ is for polarization photodetector.}
\label{F5}
\end{figure}

\textit{ii) Two controls}: Now, we will describe CPT of entangled states using two controls. This scheme is similar to that using one control as shown above. Let B$_{1}$ be the new control as shown in Fig. \ref{F5}. The atoms C and D share a previously prepared entangled state to be teleported, while B and B$_{1}$ are the two controls. At the end of this scheme for CPT the partner C of the entangled state composed by C and D will be teleported to A, with A and D becoming the entangled state. To better explaining the procedure we take advantage of the previous scheme in Subsection \textit{ii}). Assume the atoms C and D in the state given by Eq. (\ref{CD}). The photon, previously prepared in a superposed polarization state, enters the cavity A and interacts with the atom. Next, the photon is sent to interact with the atoms trapped in the cavity B and B$_{1}$, respectively. After interacting with the atom in cavity C, a Hadamard operation is applied to the atom C and the photon passes through a QWP2. Thus, 
\begin{align}
|\delta_{3}\rangle=&\frac{1}{4}\Bigg\{|00\rangle_{BB_{1}}\Big[|L0\rangle_{FC}\left(\alpha|01\rangle_{AD}-\beta|10\rangle_{AD}\right)+|L1\rangle_{FC}\left(\alpha|01\rangle_{AD}+\beta|10\rangle_{AD}\right) \notag\\
&-i|R0\rangle_{FC}(\alpha|11\rangle_{AD}+\beta|00\rangle_{AD})-i|R1\rangle_{FC}(\alpha|11\rangle_{AD}-\beta|00\rangle_{AD}) \Big]\notag\\ 
&+|01\rangle_{BB_{1}}\Big[-|L0\rangle_{FC}\left(\alpha|11\rangle_{AD}+\beta|00\rangle_{AD}\right)-|L1\rangle_{FC}\left(\alpha|11\rangle_{AD}-\beta|00\rangle_{AD}\right) \notag\\
&-i|R0\rangle_{FC}(\alpha|01\rangle_{AD}-\beta|10\rangle_{AD})-i|R1\rangle_{FC}(\alpha|01\rangle_{AD}+\beta|10\rangle_{AD}) \Big]\notag\\
&+|10\rangle_{BB_{1}}\Big[-|L0\rangle_{FC}\left(\alpha|11\rangle_{AD}+\beta|00\rangle_{AD}\right)-|L1\rangle_{FC}\left(\alpha|11\rangle_{AD}-\beta|00\rangle_{AD}\right) \notag\\
&-i|R0\rangle_{FC}(\alpha|01\rangle_{AD}-\beta|10\rangle_{AD})-i|R1\rangle_{FC}(\alpha|01\rangle_{AD}+\beta|10\rangle_{AD}) \Big]\notag\\
&-|11\rangle_{BB_{1}}\Big[|L0\rangle_{FC}\left(\alpha|01\rangle_{AD}-\beta|10\rangle_{AD}\right)+|L1\rangle_{FC}\left(\alpha|01\rangle_{AD}+\beta|10\rangle_{AD}\right) \notag\\
&-i|R0\rangle_{FC}(\alpha|11\rangle_{AD}+\beta|00\rangle_{AD})-i|R1\rangle_{FC}(\alpha|11\rangle_{AD}-\beta|00\rangle_{AD}) \Big]\Bigg\}.
\end{align}
To conclude the controlled partial teleportation of entangled states it is necessary a measurement in the state of the photon and the atom trapped in cavity C. Moreover, the state of the controls B and B$_1$ must be known by Bob. To recover a successful teleportation Bob also needs to perform the rotations given by Table \ref{T4}.

\begin{table}
\begin{center}
\begin{tabular}{|| c | c | c | c |}
\hline MAPS & CFS & ESR & AO \\ 
\hline  $|L0\rangle_{FC}$  &  $|00\rangle_{BB_{1}}$ & $\alpha|01\rangle_{AD}-\beta|10\rangle_{AD}$ & $\sigma_z\otimes\mathbb{I}$ \\
 
\hline  $|L1\rangle_{FC}$  &  $|00\rangle_{BB_{1}}$ & $\alpha|01\rangle_{AD}+\beta|10\rangle_{AD}$ & $\mathbb{I}\otimes\mathbb{I}$ \\

\hline  $|R0\rangle_{FC}$  &  $|00\rangle_{BB_{1}}$ & $\alpha|11\rangle_{AD}+\beta|00\rangle_{AD}$ & $\sigma_x\otimes\mathbb{I}$ \\

\hline  $|R1\rangle_{FC}$  &  $|00\rangle_{BB_{1}}$ & $\alpha|11\rangle_{AD}-\beta|00\rangle_{AD}$ & $\sigma_z\sigma_x\otimes\mathbb{I}$ \\ 


\hline  $|L0\rangle_{FC}$  &  $|01\rangle_{BB_{1}}$ & $\alpha|11\rangle_{AD}+\beta|00\rangle_{AD}$ & $\sigma_x\otimes\mathbb{I}$ \\
 
\hline  $|L1\rangle_{FC}$  &  $|01\rangle_{BB_{1}}$ & $\alpha|11\rangle_{AD}-\beta|00\rangle_{AD}$ & $\sigma_z\sigma_x\otimes\mathbb{I}$ \\

\hline  $|R0\rangle_{FC}$  &  $|01\rangle_{BB_{1}}$ & $\alpha|01\rangle_{AD}-\beta|10\rangle_{AD}$ & $\sigma_z\otimes\mathbb{I}$ \\

\hline  $|R1\rangle_{FC}$  &  $|01\rangle_{BB_{1}}$ & $\alpha|01\rangle_{AD}+\beta|10\rangle_{AD}$ & $\mathbb{I}\otimes\mathbb{I}$ \\


\hline  $|L0\rangle_{FC}$  &  $|10\rangle_{BB_{1}}$ & $\alpha|11\rangle_{AD}+\beta|00\rangle_{AD}$ & $\sigma_x\otimes\mathbb{I}$ \\
 
\hline  $|L1\rangle_{FC}$  &  $|10\rangle_{BB_{1}}$ & $\alpha|11\rangle_{AD}-\beta|00\rangle_{AD}$ & $\sigma_z\sigma_x\otimes\mathbb{I}$ \\

\hline  $|R0\rangle_{FC}$  &  $|10\rangle_{BB_{1}}$ & $\alpha|01\rangle_{AD}-\beta|10\rangle_{AD}$ & $\sigma_z\otimes\mathbb{I}$ \\

\hline  $|R1\rangle_{FC}$  &  $|10\rangle_{BB_{1}}$ & $\alpha|01\rangle_{AD}+\beta|10\rangle_{AD}$ & $\mathbb{I}\otimes\mathbb{I}$ \\


\hline  $|L0\rangle_{FC}$  &  $|11\rangle_{BB_{1}}$ & $\alpha|01\rangle_{AD}-\beta|10\rangle_{AD}$ & $\sigma_z\otimes\mathbb{I}$ \\
 
\hline  $|L1\rangle_{FC}$  &  $|11\rangle_{BB_{1}}$ & $\alpha|01\rangle_{AD}+\beta|10\rangle_{AD}$ & $\mathbb{I}\otimes\mathbb{I}$ \\

\hline  $|R0\rangle_{FC}$  &  $|11\rangle_{BB_{1}}$ & $\alpha|11\rangle_{AD}+\beta|00\rangle_{AD}$ & $\sigma_x\otimes\mathbb{I}$ \\

\hline  $|R1\rangle_{FC}$  &  $|11\rangle_{BB_{1}}$ & $\alpha|11\rangle_{AD}-\beta|00\rangle_{AD}$ & $\sigma_z\sigma_x\otimes\mathbb{I}$ \\ 
\hline
\end{tabular}
\end{center}
\caption{Possible results and rotations for completing the controlled partial teleportation procedure for the case \textit{ii}). The first column shows the possible results of measurements on the atom $C$ and photon $F$ states. $CFS$ and $ESR$ columns do the same for the control state ($BB_{1}$) and for the teleported state ($AD$). The fourth column  ($AO$) shows the corresponding Pauli matrices representing unitary operations upon the atomic state required to complete the teleportation process.}
\label{T4}
\end{table}

\textit{iii) Generalization}: To generalize the number of controls in thescheme one can insert the QWP after the last control as follows: QWP$_{1}$ for odd controls and QWP$_{2}$ for even controls.

\subsection{Controlled teleportation of entangled states}
\noindent

\noindent Here we present a scheme for controlled teleportation (CT) of entangled atomic states. The difference between the present scheme and that in Subsection B above, is that now the teleportation is total, \textit{%
i.e}., the teleportation scheme works for the whole entanglements of two or more particles. In this scheme the state of one of two entangled atom is rightly teleported by another atom while the atomic state remaining is teleported aided by controls. Fig. \ref{F5} displays the scheme of the CPT of entangled states, which is detailed below.

\begin{figure}[t]
\centering
\includegraphics[width=7cm]{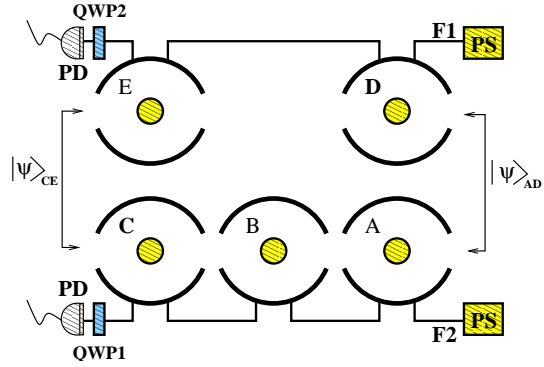}
\caption{Schematic representation for controlled teleportation of entangled
atomic states with two controls. $PS$ is for photon source, $A$, $B$, $C$, $D
$, and $E$ represent the atoms trapped in the cavities, $QWP1$ and $QWP2$
are for quarter wave plates, and $PD$ represents the polarization photodetector. $F1$
and $F2$ are for the two photonic channels.}
\label{F6}
\end{figure}

\textit{i) One control}: in this case we deal with two photons $1$ and $2$ that enters into the cavities D and A, respectively. As shown in Fig. \ref{F6}, a QWP is placed after the cavity E. The atom trapped inside the cavity B plays the role of control while the trapped atoms trapped inside the cavities C and E share the entangled state given by
\begin{equation}
|\psi\rangle_{CE}=\alpha|01\rangle_{CE}+\beta|10\rangle_{CE}.
\label{CTES00}
\end{equation} 
The other atoms and photons of the scheme are all previously prepared in a superposition of $|0\rangle$ and $|1\rangle$ or $|L\rangle$ and $|R\rangle$, respectively. The scheme also requires a QWP$_1$ after the cavity C.

Firstly we consider the photon in the inferior branch of Fig. \ref{F6}. It crosses the cavities A, B and C, with the same phase. At this point, the system can be written as  
\begin{align}
|\psi_{2}\rangle=&\frac{1}{2\sqrt{2}}\Bigg\{|0\rangle_{B}\Bigg[-\alpha|0\rangle_{A}|0\rangle_{C}|1\rangle_{E}\Big(|L\rangle_{2}+i|R\rangle_{2}\Big) + i\alpha|1\rangle_{A}|0\rangle_{C}|1\rangle_{E}\Big(|L\rangle_{2}-i|R\rangle_{2}\Big)\notag\\
&+i\beta|0\rangle_{A}|1\rangle_{C}|0\rangle_{E}\Big(|L\rangle_{2}-i|R\rangle_{2}\Big)+\beta|1\rangle_{A}|1\rangle_{C}|0\rangle_{E}\Big(|L\rangle_{2}+i|R\rangle_{2}\Big)\Bigg]+\notag\\
&|1\rangle_{B}\Bigg[i\alpha|0\rangle_{A}|0\rangle_{C}|1\rangle_{E}\Big(|L\rangle_{2}-i|R\rangle_{2}\Big)+\alpha|1\rangle_{A}|0\rangle_{C}|1\rangle_{E}\Big(|L\rangle_{2}+i|R\rangle_{2}\Big)\notag\\
&+\beta|0\rangle_{A}|1\rangle_{C}|0\rangle_{E}\Big(|L\rangle_{2}+i|R\rangle_{2}\Big)-i\beta|1\rangle_{A}|1\rangle_{C}|0\rangle_{E}\Big(|L\rangle_{2}-i|R\rangle_{2}\Big)\Bigg]\Bigg\} 
\end{align}

Now, if we consider the superior branch, the photon F1 passes through the cavities D and E, respectively. Inserting this fact in the last equation, we have
\begin{align}
|\psi_{2}\rangle=&\frac{1}{4\sqrt{2}}\Bigg\{|0\rangle_{B}\Bigg[-\alpha|0\rangle_{A}|0\rangle_{C}|1\rangle_{E}\Big(|L\rangle_{2}+i|R\rangle_{2}\Big)\Big[-i\big(|L\rangle_{1}+|R\rangle_{1}\Big)|0\rangle_{D}-\Big(|L\rangle_{1}-|R\rangle_{1}\Big)|1\rangle_{D}  \Big]\notag\\
&+i\alpha|1\rangle_{A}|0\rangle_{C}|1\rangle_{E}\Big(|L\rangle_{2}-i|R\rangle_{2}\Big)\Big[-i\big(|L\rangle_{1}+|R\rangle_{1}\Big)|0\rangle_{D}-\Big(|L\rangle_{1}-|R\rangle_{1}\Big)|1\rangle_{D}\Big]\notag\\
&+i\beta|0\rangle_{A}|1\rangle_{C}|0\rangle_{E}\Big(|L\rangle_{2}-i|R\rangle_{2}\Big)\Big[\Big(|L\rangle_{1}-|R\rangle_{1}\Big)|0\rangle_{D}-i\Big(|L\rangle_{1}+|R\rangle_{1}\Big)|1\rangle_{D}\Big]\notag\\
&+\beta|1\rangle_{A}|1\rangle_{C}|0\rangle_{E}\Big(|L\rangle_{2}+i|R\rangle_{2}\Big)\Big[\Big(|L\rangle_{1}-|R\rangle_{1}\Big)|0\rangle_{D}-i\Big(|L\rangle_{1}+|R\rangle_{1}\Big)|1\rangle_{D}\Big]\Bigg]+\notag\\
&|1\rangle_{B}\Bigg[i\alpha|0\rangle_{A}|0\rangle_{C}|1\rangle_{E}\Big(|L\rangle_{2}-i|R\rangle_{2}\Big)\Big[-i\Big(|L\rangle_{1}+|R\rangle_{1}\Big)|0\rangle_{D}-\Big(|L\rangle_{1}-|R\rangle_{1}\Big)|1\rangle_{D}  \Big]\notag\\
&+\alpha|1\rangle_{A}|0\rangle_{C}|1\rangle_{E}\Big(|L\rangle_{2}+i|R\rangle_{2}\Big)\Big[-i\Big(|L\rangle_{1}+|R\rangle_{1}\Big)|0\rangle_{D}-\Big(|L\rangle_{1}-|R\rangle_{1}\Big)|1\rangle_{D}\Big]\notag\\
&+\beta|0\rangle_{A}|1\rangle_{C}|0\rangle_{E}\Big(|L\rangle_{2}+i|R\rangle_{2}\Big)\Big[\Big(|L\rangle_{1}-|R\rangle_{1}\Big)|0\rangle_{D}-i\Big(|L\rangle_{1}+|R\rangle_{1}\Big)|1\rangle_{D}\Big]\notag\\
&-i\beta|1\rangle_{A}|1\rangle_{C}|0\rangle_{E}\Big(|L\rangle_{2}-i|R\rangle_{2}\Big)\Big[\Big(|L\rangle_{1}-|R\rangle_{1}\Big)|0\rangle_{D}-i\Big(|L\rangle_{1}+|R\rangle_{1}\Big)|1\rangle_{D}\Big]\Bigg]\Bigg\} 
\end{align}

After, the photons F1 and F2 suffer a Hadamard operation by the plates QWP2 and QWP1, respectively. Besides, other Hadamard operations in the atoms E and C lead the system to
\begin{align}
|\psi_{3}\rangle=&\frac{\sqrt{2}}{4}\Bigg\{|0\rangle_{B}\Bigg[|R0\rangle_{2C}|L0\rangle_{1E}\big(\alpha|10\rangle_{AD}+\beta|01\rangle_{AD}\big) + |R1\rangle_{2C}|L0\rangle_{1E}\big(\alpha|10\rangle_{AD}-\beta|01\rangle_{AD}\big)\notag\\
&-|R0\rangle_{2C}|L1\rangle_{1E}\big(\alpha|10\rangle_{AD}-\beta|01\rangle_{AD}\big) - |R1\rangle_{2C}|L1\rangle_{1E}\big(\alpha|10\rangle_{AD}+\beta|01\rangle_{AD}\big)\notag\\
&-i|R0\rangle_{2C}|R0\rangle_{1E}\big(\alpha|11\rangle_{AD}-\beta|00\rangle_{AD}\big) + i|R0\rangle_{2C}|R1\rangle_{1E}\big(\alpha|11\rangle_{AD}+\beta|00\rangle_{AD}\big)\notag\\
&-i|R1\rangle_{2C}|R0\rangle_{1E}\big(\alpha|11\rangle_{AD}+\beta|00\rangle_{AD}\big) + i|R1\rangle_{2C}|R1\rangle_{1E}\big(\alpha|11\rangle_{AD}-\beta|00\rangle_{AD}\big)\notag\\
&+i|L0\rangle_{2C}|L0\rangle_{1E}\big(\alpha|00\rangle_{AD}-\beta|11\rangle_{AD}\big) - i|L0\rangle_{2C}|L1\rangle_{1E}\big(\alpha|00\rangle_{AD}+\beta|11\rangle_{AD}\big)\notag\\
&+i|L1\rangle_{2C}|L0\rangle_{1E}\big(\alpha|00\rangle_{AD}+\beta|11\rangle_{AD}\big) - i|L1\rangle_{2C}|L1\rangle_{1E}\big(\alpha|00\rangle_{AD}-\beta|11\rangle_{AD}\big)\notag\\
&+|L0\rangle_{2C}|R0\rangle_{1E}\big(\alpha|01\rangle_{AD}+\beta|10\rangle_{AD}\big) - |L0\rangle_{2C}|R1\rangle_{1E}\big(\alpha|01\rangle_{AD}-\beta|10\rangle_{AD}\big)\notag\\
&+|L1\rangle_{2C}|R0\rangle_{1E}\big(\alpha|01\rangle_{AD}-\beta|10\rangle_{AD}\big) - |L1\rangle_{2C}|R1\rangle_{1E}\big(\alpha|01\rangle_{AD}+\beta|10\rangle_{AD}\big)\Bigg]+\notag\\
&|1\rangle_{B}\Bigg[|R0\rangle_{2C}|L0\rangle_{1E}\big(\alpha|00\rangle_{AD}-\beta|11\rangle_{AD}\big)-  |R0\rangle_{2C}|L1\rangle_{1E}(\alpha|00\rangle_{AD}+\beta|11\rangle_{AD})\notag\\
&+|R1\rangle_{2C}|L0\rangle_{1E}\big(\alpha|00\rangle_{AD}+\beta|11\rangle_{AD}\big) - |R1\rangle_{2C}|L1\rangle_{1E}\big(\alpha|00\rangle_{AD}-\beta|11\rangle_{AD}\big)\notag\\
&-i|R0\rangle_{2C}|R0\rangle_{1E}\big(\alpha|01\rangle_{AD}+\beta|10\rangle_{AD}\big) +i|R0\rangle_{2C}|R1\rangle_{1E}\big(\alpha|01\rangle_{AD}-\beta|10\rangle_{AD}\big)\notag\\
&-i|R1\rangle_{2C}|R0\rangle_{1E}\big(\alpha|01\rangle_{AD}-\beta|10\rangle_{AD}\big) + i|R1\rangle_{2C}|R1\rangle_{1E}\big(\alpha|01\rangle_{AD}+\beta|10\rangle_{AD}\big)\notag\\
&-i|L0\rangle_{2C}|L0\rangle_{1E}\big(\alpha|10\rangle_{AD}+\beta|01\rangle_{AD}\big) + i|L0\rangle_{2C}|L1\rangle_{1E}\big(\alpha|10\rangle_{AD}-\beta|01\rangle_{AD}\big)\notag\\
&-i|L1\rangle_{2C}|R0\rangle_{1E}\big(\alpha|10\rangle_{AD}-\beta|01\rangle_{AD}\big) + i|L1\rangle_{2C}|R1\rangle_{1E}\big(\alpha|10\rangle_{AD}+\beta|01\rangle_{AD}\big)\notag\\
&-|L0\rangle_{2C}|R0\rangle_{1E}\big(\alpha|11\rangle_{AD}-\beta|00\rangle_{AD}\big) + |L0\rangle_{2C}|R1\rangle_{1E}\big(\alpha|11\rangle_{AD}+\beta|00\rangle_{AD}\big)\notag\\
&-|L1\rangle_{2C}|R0\rangle_{1E}\big(\alpha|11\rangle_{AD}+\beta|00\rangle_{AD}\big) + |L0\rangle_{2C}|R1\rangle_{1E}\big(\alpha|11\rangle_{AD}-\beta|00\rangle_{AD}\big)\Bigg]\Bigg\}
\end{align}

At this time the measurement can be concluded with a single measurement in the state of the photons 1 and 2 as well as those of the atoms C and E. The teleportation will be finalized after the knowledge of these results plus the results of the atomic state of the control (B). The probable results are summarized in the Table \ref{T5}.

\begin{table}
\begin{center}
\begin{tabular}{|| c | c | c | c | c |}
\hline 2C & 1$E$ & Control & $AD$ & AO \\ 
\hline  $|R0\rangle_{2C}$  & $|L0\rangle_{1E}$ & $|0\rangle_{B}$ & $\alpha|10\rangle_{AD}+\beta|01\rangle_{AD}$ & $\sigma_x\otimes\sigma_x$ \\
 
\hline  $|R1\rangle_{2C}$  & $|L1\rangle_{1E}$ & $|0\rangle_{B}$ & $\alpha|10\rangle_{AD}-\beta|01\rangle_{AD}$ & $\sigma_z\sigma_x\otimes\sigma_x$ \\

\hline  $|R0\rangle_{2C}$  & $|L1\rangle_{1E}$ & $|0\rangle_{B}$ & $\alpha|10\rangle_{AD}-\beta|01\rangle_{AD}$ & $\sigma_z\sigma_x\otimes\sigma_x$ \\

\hline  $|R1\rangle_{2C}$  & $|L1\rangle_{1E}$ & $|0\rangle_{B}$ & $\alpha|10\rangle_{AD}+\beta|01\rangle_{AD}$ & $\sigma_x\otimes\sigma_x$ \\
	

\hline  $|R0\rangle_{2C}$  & $|R0\rangle_{1E}$ & $|0\rangle_{B}$ & $\alpha|11\rangle_{AD}-\beta|00\rangle_{AD}$ & $\sigma_z\sigma_x\otimes\mathbb{I}$ \\
 
\hline  $|R0\rangle_{2C}$  & $|R1\rangle_{1E}$ & $|0\rangle_{B}$ & $\alpha|11\rangle_{AD}+\beta|00\rangle_{AD}$ & $\sigma_x\otimes\mathbb{I}$ \\

\hline  $|R1\rangle_{2C}$  & $|R0\rangle_{1E}$ & $|0\rangle_{B}$ & $\alpha|11\rangle_{AD}+\beta|00\rangle_{AD}$ & $\sigma_x\otimes\mathbb{I}$ \\

\hline  $|R1\rangle_{2C}$  & $|R1\rangle_{1E}$ & $|0\rangle_{B}$ & $\alpha|11\rangle_{AD}-\beta|00\rangle_{AD}$ & $\sigma_z\sigma_x\otimes\mathbb{I}$ \\


\hline  $|L0\rangle_{2C}$  & $|L0\rangle_{1E}$ & $|0\rangle_{B}$ & $\alpha|00\rangle_{AD}-\beta|11\rangle_{AD}$ & $\mathbb{I}\otimes\sigma_x\sigma_z$ \\

\hline  $|L0\rangle_{2C}$  & $|L1\rangle_{1E}$ & $|0\rangle_{B}$ & $\alpha|00\rangle_{AD}+\beta|11\rangle_{AD}$ & $\mathbb{I}\otimes\sigma_x$ \\

\hline  $|L1\rangle_{2C}$  & $|L0\rangle_{1E}$ & $|0\rangle_{B}$ & $\alpha|00\rangle_{AD}+\beta|11\rangle_{AD}$ & $\mathbb{I}\otimes\sigma_x$ \\

\hline  $|L1\rangle_{2C}$  & $|L1\rangle_{1E}$ & $|0\rangle_{B}$ & $\alpha|00\rangle_{AD}-\beta|11\rangle_{AD}$ & $\mathbb{I}\otimes\sigma_x\sigma_z$ \\


\hline  $|L0\rangle_{2C}$  & $|R0\rangle_{1E}$ & $|0\rangle_{B}$ & $\alpha|01\rangle_{AD}+\beta|10\rangle_{AD}$ & $\mathbb{I}\otimes\mathbb{I}$ \\
 
\hline  $|L0\rangle_{2C}$  & $|R1\rangle_{1E}$ & $|0\rangle_{B}$ & $\alpha|01\rangle_{AD}-\beta|10\rangle_{AD}$ & $\sigma_z\otimes\mathbb{I}$ \\

\hline  $|L1\rangle_{2C}$  & $|R0\rangle_{1E}$ & $|0\rangle_{B}$ & $\alpha|01\rangle_{AD}-\beta|10\rangle_{AD}$ & $\sigma_z\otimes\mathbb{I}$ \\

\hline  $|L1\rangle_{2C}$  & $|R1\rangle_{1E}$ & $|0\rangle_{B}$ & $\alpha|01\rangle_{AD}+\beta|10\rangle_{AD}$ & $\mathbb{I}\otimes\mathbb{I}$ \\
\hline
\end{tabular}
\begin{tabular}{|| c | c | c | c | c |}
\hline 2C & 1$E$ & Control & $AD$ & AO \\ 
\hline  $|R0\rangle_{2C}$  & $|L0\rangle_{1E}$ & $|1\rangle_{B}$ & $\alpha|00\rangle_{AD}-\beta|11\rangle_{AD}$ & $\mathbb{I}\otimes\sigma_x\sigma_z$ \\
 
\hline  $|R0\rangle_{2C}$  & $|L1\rangle_{1E}$ & $|1\rangle_{B}$ & $\alpha|00\rangle_{AD}+\beta|11\rangle_{AD}$ & $\mathbb{I}\otimes\sigma_x$ \\

\hline  $|R1\rangle_{2C}$  & $|L0\rangle_{1E}$ & $|1\rangle_{B}$ & $\alpha|00\rangle_{AD}+\beta|11\rangle_{AD}$ & $\mathbb{I}\otimes\sigma_x$ \\

\hline  $|R1\rangle_{2C}$  & $|L1\rangle_{1E}$ & $|1\rangle_{B}$ & $\alpha|00\rangle_{AD}-\beta|11\rangle_{AD}$ & $\mathbb{I}\otimes\sigma_x\sigma_z$ \\
	

\hline  $|R0\rangle_{2C}$  & $|R0\rangle_{1E}$ & $|1\rangle_{B}$ & $\alpha|01\rangle_{AD}+\beta|10\rangle_{AD}$ & $\mathbb{I}\otimes\mathbb{I}$ \\
 
\hline  $|R0\rangle_{2C}$  & $|R1\rangle_{1E}$ & $|1\rangle_{B}$ & $\alpha|01\rangle_{AD}-\beta|10\rangle_{AD}$ & $\sigma_z\otimes\mathbb{I}$ \\

\hline  $|R1\rangle_{2C}$  & $|R0\rangle_{1E}$ & $|1\rangle_{B}$ & $\alpha|01\rangle_{AD}-\beta|10\rangle_{AD}$ & $\sigma_z\otimes\mathbb{I}$ \\

\hline  $|R1\rangle_{2C}$  & $|R1\rangle_{1E}$ & $|1\rangle_{B}$ & $\alpha|01\rangle_{AD}+\beta|10\rangle_{AD}$ & $\mathbb{I}\otimes\mathbb{I}$ \\


\hline  $|L0\rangle_{2C}$  & $|L0\rangle_{1D}$ & $|1\rangle_{B}$ & $\alpha|10\rangle_{AD}+\beta|01\rangle_{AD}$ & $\sigma_x\otimes\sigma_x$ \\

\hline  $|L0\rangle_{2C}$  & $|L1\rangle_{1E}$ & $|1\rangle_{B}$ & $\alpha|10\rangle_{AD}-\beta|01\rangle_{AD}$ & $\sigma_z\sigma_x\otimes\sigma_x$ \\

\hline  $|L1\rangle_{2C}$  & $|L0\rangle_{1C_{1}}$ & $|1\rangle_{B}$ & $\alpha|10\rangle_{AA_{1}}-\beta|01\rangle_{AA_{1}}$ & $\sigma_z\sigma_x\otimes\sigma_x$ \\

\hline  $|L1\rangle_{2C}$  & $L1\rangle_{1E}$ & $|1\rangle_{B}$ & $\alpha|10\rangle_{AD}+\beta|01\rangle_{AD}$ & $\sigma_x\otimes\sigma_x$ \\


\hline  $|L0\rangle_{2C}$  & $|R0\rangle_{1E}$ & $|1\rangle_{B}$ & $\alpha|11\rangle_{AD}-\beta|00\rangle_{AD}$ & $\sigma_z\sigma_x\otimes\mathbb{I}$ \\
 
\hline  $|L0\rangle_{2C}$  & $|R1\rangle_{1E}$ & $|1\rangle_{B}$ & $\alpha|11\rangle_{AD}+\beta|00\rangle_{AD}$ & $\sigma_x\otimes\mathbb{I}$ \\

\hline  $|L1\rangle_{2C}$  & $|R0\rangle_{1E}$ & $|1\rangle_{B}$ & $\alpha|11\rangle_{AD}+\beta|00\rangle_{AD}$ & $\sigma_x\otimes\mathbb{I}$ \\

\hline  $|L1\rangle_{2C}$  & $|R1\rangle_{1E}$ & $|1\rangle_{B}$ & $\alpha|11\rangle_{AD}-\beta|00\rangle_{AD}$ & $\sigma_z\sigma_x\otimes\mathbb{I}$ \\
\hline 
\end{tabular}
\end{center}
\caption{Possible results and rotations for completing the controlled teleportation with one control, corresponding to case \textit{i}). The first column (2C) shows the possible results of measurements on the states of atom $C$ and photon $2$. Second colunm refers to the Alice's possible results of measurements on photon $1$ atom $E$. Column $Control$ shows the state of control B.  $AD$  column does the the same for the teleported state ($AD$). Fourth column ($AO$) shows the corresponding Pauli matrices representing unitary operations on the atomic state required to complete the teleportation process.}
\label{T5}
\end{table}

\begin{figure}[t]
\centering
\includegraphics[width=8cm]{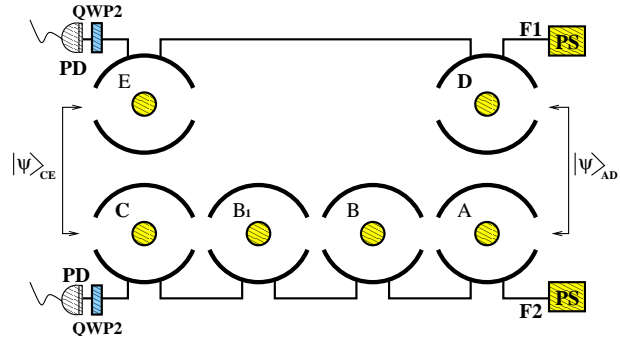}
\caption{Schematic representation for controlled teleportation of entangled atomic states with two controls. $PS$ is for photon sources, $A$, $B$, $B_{1} $, $C$, $D$, and $E$ represent the atoms trapped in the cavities, $QWP2$ is
for the two identical quarter wave plates, and $PD$ is for polarization photodetector. $F1$ and $F2$ are for the two photonic channels.}\label{F7}
\end{figure}

\textit{ii) Two controls}: In this case we consider an additional cavity B$_{1}$ with the trapped atom working as the new control of the teleportation procedure. The scheme is illustrated in the Fig. \ref{F7}. Note that the upper path is not modified. On the other hand, the state of the subsystem of the lower branch, after the interaction of the photon 2 with the atoms trapped in the cavities A, B, B$_{1}$, and C, respectively, is given by

\begin{align}
|\phi_{3}^{'}\rangle=&\frac{1}{4}\Big(|00\rangle_{B B_{1}}\Big[\alpha\big(|L\rangle_{2} + |R\rangle_{2}\big)|0\rangle_{A}|01\rangle_{CE} -i\alpha\big(|L\rangle_{2} - |R\rangle_{2}\big)|1\rangle_{A}|01\rangle_{CE}\notag\\
&-i\beta\big(|L\rangle_{2} - |R\rangle_{2}\big)|0\rangle_{A}|10\rangle_{CE}-\beta(|L\rangle_{2} + |R\rangle_{2})|1\rangle_{A}|10\rangle_{CE}  \Big]\notag\\
&+|01\rangle_{B B_{1}}\Big[-i\alpha\big(|L\rangle_{2} - |R\rangle_{2}\big)|0\rangle_{A}|01\rangle_{CE} - \alpha\big(|L\rangle_{2} + |R\rangle_{2}\big)|1\rangle_{A}|01\rangle_{CE}\notag\\
&-\beta\big(|L\rangle_{2} + |R\rangle_{2}\big)|0\rangle_{A}|10\rangle_{CE}+i\beta\big(|L\rangle_{2} - |R\rangle_{2}\big)|1\rangle_{A}|10\rangle_{CE}  \Big]\notag\\
&+|10\rangle_{B B_{1}}\Big[-i\alpha\big(|L\rangle_{2} - |R\rangle_{2}\big)|0\rangle_{A}|01\rangle_{CE} - \alpha\big(|L\rangle_{2} + |R\rangle_{2}\big)|1\rangle_{A}|01\rangle_{CE}\notag\\
&-\beta\big(|L\rangle_{2} + |R\rangle_{2}\big)|0\rangle_{A}|10\rangle_{CE}+i\beta\big(|L\rangle_{2} - |R\rangle_{2}\big)|1\rangle_{A}|10\rangle_{CE}  \Big]\notag\\
&-|11\rangle_{B B_{1}}\Big[\alpha\big(|L\rangle_{2} + |R\rangle_{2}\big)|0\rangle_{A}|01\rangle_{CE} -i\alpha\big(|L\rangle_{2} - |R\rangle_{2}\big)|1\rangle_{A}|01\rangle_{CE}\notag\\
&-i\beta\big(|L\rangle_{2} - |R\rangle_{2}\big)|0\rangle_{A}|10\rangle_{CE}-\beta\big(|L\rangle_{2} + |R\rangle_{2}\big)|1\rangle_{A}|10\rangle_{CE}  \Big]\Big). 
\end{align}

Now, after the photon F1 has interacted with D and E, the state of the system can be written as
\begin{align}
|\phi_{3}^{''}\rangle=&\frac{1}{8}\Bigg\{|00\rangle_{B B_{1}}\Bigg[\alpha\big(|L\rangle_{2} + |R\rangle_{2}\big)|0\rangle_{A}|01\rangle_{CE}\Big[-i\Big(|L\rangle_{1}+|R\rangle_{1}\Big)|0\rangle_{D}- \Big(|L\rangle_{1}-|R\rangle_{1}\Big)|1\rangle_{D}   \Big]\notag\\
&-i\alpha\big(|L\rangle_{2} - |R\rangle_{2}\big)|1\rangle_{A}|01\rangle_{CE}\Big[-i\Big(|L\rangle_{1}+|R\rangle_{1}\Big)|0\rangle_{D}- \Big(|L\rangle_{1}-|R\rangle_{1}\Big)|1\rangle_{D}   \Big]\notag\\
&-i\beta\big(|L\rangle_{2} - |R\rangle_{2}\big)|0\rangle_{A}|10\rangle_{CE}\Big[\Big(|L\rangle_{1}-|R\rangle_{1}\Big)|0\rangle_{D}- i\Big(|L\rangle_{1}+|R\rangle_{1}\Big)|1\rangle_{D}   \Big]\notag\\
&-\beta(|L\rangle_{2} + |R\rangle_{2})|1\rangle_{A}|10\rangle_{CE}\Big[\Big(|L\rangle_{1}-|R\rangle_{1}\Big)|0\rangle_{D}- i\Big(|L\rangle_{1}+|R\rangle_{1}\Big)|1\rangle_{D}   \Big]  \Bigg]\notag\\
&+|01\rangle_{B B_{1}}\Bigg[-i\alpha\big(|L\rangle_{2} - |R\rangle_{2}\big)|0\rangle_{A}|01\rangle_{CE}\Big[-i\Big(|L\rangle_{1}+|R\rangle_{1}\Big)|0\rangle_{D}- \Big(|L\rangle_{1}-|R\rangle_{1}\Big)|1\rangle_{D}   \Big]\notag\\
&- \alpha\big(|L\rangle_{2} + |R\rangle_{2}\big)|1\rangle_{A}|01\rangle_{CE}\Big[-i\Big(|L\rangle_{1}+|R\rangle_{1}\Big)|0\rangle_{D}- \Big(|L\rangle_{1}-|R\rangle_{1}\Big)|1\rangle_{D}   \Big]\notag\\
&-\beta\big(|L\rangle_{2} + |R\rangle_{2}\big)|0\rangle_{A}|10\rangle_{CE}\Big[\Big(|L\rangle_{1}-|R\rangle_{1}\Big)|0\rangle_{D}- i\Big(|L\rangle_{1}+|R\rangle_{1}\Big)|1\rangle_{D}   \Big]\notag\\
&+i\beta\big(|L\rangle_{2} - |R\rangle_{2}\big)|1\rangle_{A}|10\rangle_{CE}\Big[\Big(|L\rangle_{1}-|R\rangle_{1}\Big)|0\rangle_{D}- i\Big(|L\rangle_{1}+|R\rangle_{1}\Big)|1\rangle_{D}   \Big]  \Bigg]\notag\\
&+|10\rangle_{B B_{1}}\Bigg[-i\alpha\big(|L\rangle_{2} - |R\rangle_{2}\big)|0\rangle_{A}|01\rangle_{CE}\Big[-i\Big(|L\rangle_{1}+|R\rangle_{1}\Big)|0\rangle_{D}- \Big(|L\rangle_{1}-|R\rangle_{1}\Big)|1\rangle_{D}   \Big]\notag\\
& - \alpha\big(|L\rangle_{2} + |R\rangle_{2}\big)|1\rangle_{A}|01\rangle_{CE}\Big[-i\Big(|L\rangle_{1}+|R\rangle_{1}\Big)|0\rangle_{D}- \Big(|L\rangle_{1}-|R\rangle_{1}\Big)|1\rangle_{D}   \Big]\notag\\
&-\beta\big(|L\rangle_{2} + |R\rangle_{2}\big)|0\rangle_{A}|10\rangle_{CE}\Big[\Big(|L\rangle_{1}-|R\rangle_{1}\Big)|0\rangle_{D}- i\Big(|L\rangle_{1}+|R\rangle_{1}\Big)|1\rangle_{D}   \Big]\notag\\
&+i\beta\big(|L\rangle_{2} - |R\rangle_{2}\big)|1\rangle_{A}|10\rangle_{CE}\Big[\Big(|L\rangle_{1}-|R\rangle_{1}\Big)|0\rangle_{D}- i\Big(|L\rangle_{1}+|R\rangle_{1}\Big)|1\rangle_{D}   \Big]  \Bigg]\notag\\
&-|11\rangle_{B B_{1}}\Bigg[\alpha\big(|L\rangle_{2} + |R\rangle_{2}\big)|0\rangle_{A}|01\rangle_{CE}\Big[-i\Big(|L\rangle_{1}+|R\rangle_{1}\Big)|0\rangle_{D}- \Big(|L\rangle_{1}-|R\rangle_{1}\Big)|1\rangle_{D}   \Big]\notag\\
& -i\alpha\big(|L\rangle_{2} - |R\rangle_{2}\big)|1\rangle_{A}|01\rangle_{CE}\Big[-i\Big(|L\rangle_{1}+|R\rangle_{1}\Big)|0\rangle_{D}- \Big(|L\rangle_{1}-|R\rangle_{1}\Big)|1\rangle_{D}   \Big]\notag\\
&-i\beta\big(|L\rangle_{2} - |R\rangle_{2}\big)|0\rangle_{A}|10\rangle_{CE}\Big[\Big(|L\rangle_{1}-|R\rangle_{1}\Big)|0\rangle_{D}- i\Big(|L\rangle_{1}+|R\rangle_{1}\Big)|1\rangle_{D}   \Big]\notag\\
&-\beta\big(|L\rangle_{2} + |R\rangle_{2}\big)|1\rangle_{A}|10\rangle_{CE}\Big[\Big(|L\rangle_{1}-|R\rangle_{1}\Big)|0\rangle_{D}- i\Big(|L\rangle_{1}+|R\rangle_{1}\Big)|1\rangle_{D}   \Big]  \Bigg]\Bigg\}. 
\end{align}

Next, considering the plate QWP2 for F1 and F2, plus a Hadamard operations upon C and E, we obtain
\begin{align}
|\psi_{3}\rangle=&\frac{1}{8}\Bigg\{|00\rangle_{BB_{1}}\Bigg[-i|L0\rangle_{2C}|L0\rangle_{1E}(\alpha|00\rangle_{AD}-\beta|11\rangle_{AD}) + i|L0\rangle_{2C}|L1\rangle_{1E}(\alpha|00\rangle_{AD}+\beta|11\rangle_{AD})\notag\\
&-i|L1\rangle_{2C}|L0\rangle_{1E}(\alpha|00\rangle_{AD}+\beta|11\rangle_{AD}) + i|L1\rangle_{2C}|L1\rangle_{1E}(\alpha|00\rangle_{AD}-\beta|11\rangle_{AD})\notag\\
&-|L0\rangle_{2C}|R0\rangle_{1E}(\alpha|01\rangle_{AD}+\beta|10\rangle_{AD}) + |L0\rangle_{2C}|R1\rangle_{1E}(\alpha|01\rangle_{AD}-\beta|10\rangle_{AD})\notag\\
&-|L1\rangle_{2C}|R0\rangle_{1E}(\alpha|01\rangle_{AD}-\beta|10\rangle_{AD}) + |L1\rangle_{2C}|R1\rangle_{1E}(\alpha|01\rangle_{AD}+\beta|10\rangle_{AD})\notag\\
&-|R0\rangle_{2C}|L0\rangle_{1E}(\alpha|10\rangle_{AD}+\beta|01\rangle_{AD}) + |R0\rangle_{2C}|L1\rangle_{1E}(\alpha|10\rangle_{AD}-\beta|01\rangle_{AD})\notag\\
&-|R1\rangle_{2C}|L0\rangle_{1E}(\alpha|10\rangle_{AD}-\beta|01\rangle_{AD}) + |R1\rangle_{2C}|L1\rangle_{1E}(\alpha|10\rangle_{AD}+\beta|01\rangle_{AD})\notag\\
&+i|R0\rangle_{2C}|R0\rangle_{1E}(\alpha|11\rangle_{AD}-\beta|00\rangle_{AD}) - i|R0\rangle_{2C}|R1\rangle_{1E}(\alpha|11\rangle_{AD}+\beta|00\rangle_{AD})\notag\\
&+i|R1\rangle_{2C}|R0\rangle_{1E}(\alpha|11\rangle_{AD}+\beta|00\rangle_{AD}) - i|R1\rangle_{2C}|R1\rangle_{1E}(\alpha|11\rangle_{AD}-\beta|00\rangle_{AD})\Bigg]\notag\\
&+|01\rangle_{BB_{1}}\Bigg[i|L0\rangle_{2C}|L0\rangle_{1E}(\alpha|10\rangle_{AD}+\beta|01\rangle_{AD}) - i|L0\rangle_{2C}|L1\rangle_{1E}(\alpha|10\rangle_{AD}-\beta|01\rangle_{AD})\notag\\
&+i|L1\rangle_{2C}|L0\rangle_{1E}(\alpha|10\rangle_{AD}-\beta|01\rangle_{AD}) - i|L1\rangle_{2C}|L1\rangle_{1E}(\alpha|10\rangle_{AD}+\beta|01\rangle_{AD})\notag\\
&+|L0\rangle_{2C}|R0\rangle_{1E}(\alpha|11\rangle_{AD}-\beta|00\rangle_{AD}) - |L0\rangle_{2C}|R1\rangle_{1E}(\alpha|11\rangle_{AD}+\beta|00\rangle_{AD})\notag\\
&+|L1\rangle_{2C}|R0\rangle_{1E}(\alpha|11\rangle_{AD}+\beta|00\rangle_{AD}) - |L1\rangle_{2C}|R1\rangle_{1E}(\alpha|11\rangle_{AD}-\beta|00\rangle_{AD})\notag\\
&-|R0\rangle_{2C}|L0\rangle_{1E}(\alpha|00\rangle_{AD}-\beta|11\rangle_{AD}) + |R0\rangle_{2C}|L1\rangle_{1E}(\alpha|00\rangle_{AD}+\beta|11\rangle_{AD})\notag\\
&-|R1\rangle_{2C}|L0\rangle_{1E}(\alpha|00\rangle_{AD}+\beta|11\rangle_{AD}) + |R1\rangle_{2C}|L1\rangle_{1E}(\alpha|00\rangle_{AD}-\beta|11\rangle_{AD})\notag\\
&+i|R0\rangle_{2C}|R0\rangle_{1E}(\alpha|01\rangle_{AD}+\beta|10\rangle_{AD}) - i|R0\rangle_{2C}|R1\rangle_{1E}(\alpha|01\rangle_{AD}-\beta|10\rangle_{AD})\notag\\
&+i|R1\rangle_{2C}|R0\rangle_{1E}(\alpha|01\rangle_{AD}-\beta|10\rangle_{AD}) - i|R1\rangle_{2C}|R1\rangle_{1E}(\alpha|01\rangle_{AD}+\beta|10\rangle_{AD})\Bigg]\notag\\
&+|10\rangle_{BB_{1}}\Bigg[i|L0\rangle_{2C}|L0\rangle_{1E}(\alpha|10\rangle_{AD}+\beta|01\rangle_{AD}) - i|L0\rangle_{2C}|L1\rangle_{1E}(\alpha|10\rangle_{AD}-\beta|01\rangle_{AD})\notag\\
&+i|L1\rangle_{2C}|L0\rangle_{1E}(\alpha|10\rangle_{AD}-\beta|01\rangle_{AD}) - i|L1\rangle_{2C}|L1\rangle_{1E}(\alpha|10\rangle_{AD}+\beta|01\rangle_{AD})\notag\\
&+|L0\rangle_{2C}|R0\rangle_{1E}(\alpha|11\rangle_{AD}-\beta|00\rangle_{AD}) - |L0\rangle_{2C}|R1\rangle_{1E}(\alpha|11\rangle_{AD}+\beta|00\rangle_{AD})\notag\\
&+|L1\rangle_{2C}|R0\rangle_{1E}(\alpha|11\rangle_{AD}+\beta|00\rangle_{AD}) - |L1\rangle_{2C}|R1\rangle_{1E}(\alpha|11\rangle_{AD}-\beta|00\rangle_{AD})\notag\\
&-|R0\rangle_{2C}|L0\rangle_{1E}(\alpha|00\rangle_{AD}-\beta|11\rangle_{AD}) + |R0\rangle_{2C}|L1\rangle_{1E}(\alpha|00\rangle_{AD}+\beta|11\rangle_{AD})\notag\\
&-|R1\rangle_{2C}|L0\rangle_{1E}(\alpha|00\rangle_{AD}+\beta|11\rangle_{AD}) + |R1\rangle_{2C}|L1\rangle_{1E}(\alpha|00\rangle_{AD}-\beta|11\rangle_{AD})\notag\\
&+i|R0\rangle_{2C}|R0\rangle_{1E}(\alpha|01\rangle_{AD}+\beta|10\rangle_{AD}) - i|R0\rangle_{2C}|R1\rangle_{1E}(\alpha|01\rangle_{AD}-\beta|10\rangle_{AD})\notag\\
&+i|R1\rangle_{2C}|R0\rangle_{1E}(\alpha|01\rangle_{AD}-\beta|10\rangle_{AD}) - i|R1\rangle_{2C}|R1\rangle_{1E}(\alpha|01\rangle_{AD}+\beta|10\rangle_{AD})\Bigg]\notag\\
&-|11\rangle_{BB_{1}}\Bigg[-i|L0\rangle_{2C}|L0\rangle_{1E}(\alpha|00\rangle_{AD}-\beta|11\rangle_{AD}) + i|L0\rangle_{2C}|L1\rangle_{1E}(\alpha|00\rangle_{AD}+\beta|11\rangle_{AD})\notag\\
&-i|L1\rangle_{2C}|L0\rangle_{1E}(\alpha|00\rangle_{AD}+\beta|11\rangle_{AD}) + i|L1\rangle_{2C}|L1\rangle_{1E}(\alpha|00\rangle_{AD}-\beta|11\rangle_{AD})\notag\\
&-|L0\rangle_{2C}|R0\rangle_{1E}(\alpha|01\rangle_{AD}+\beta|10\rangle_{AD}) + |L0\rangle_{2C}|R1\rangle_{1E}(\alpha|01\rangle_{AD}-\beta|10\rangle_{AD})\notag\\
&-|L1\rangle_{2C}|R0\rangle_{1E}(\alpha|01\rangle_{AD}-\beta|10\rangle_{AD}) + |L1\rangle_{2C}|R1\rangle_{1E}(\alpha|01\rangle_{AD}+\beta|10\rangle_{AD})\notag\\
&-|R0\rangle_{2C}|L0\rangle_{1E}(\alpha|10\rangle_{AD}+\beta|01\rangle_{AD}) + |R0\rangle_{2C}|L1\rangle_{1E}(\alpha|10\rangle_{AD}-\beta|01\rangle_{AD})\notag\\
&-|R1\rangle_{2C}|L0\rangle_{1E}(\alpha|10\rangle_{AD}-\beta|01\rangle_{AD}) + |R1\rangle_{2C}|L1\rangle_{1E}(\alpha|10\rangle_{AD}+\beta|01\rangle_{AD})\notag\\
&+i|R0\rangle_{2C}|R0\rangle_{1E}(\alpha|11\rangle_{AD}-\beta|00\rangle_{AD}) - i|R0\rangle_{2C}|R1\rangle_{1E}(\alpha|11\rangle_{AD}+\beta|00\rangle_{AD})\notag\\
&+i|R1\rangle_{2C}|R0\rangle_{1E}(\alpha|11\rangle_{AD}+\beta|00\rangle_{AD}) - i|R1\rangle_{2C}|R1\rangle_{1E}(\alpha|11\rangle_{AD}-\beta|00\rangle_{AD})\Bigg]\Bigg\}.
\end{align}
To conclude the controlled teleportation of entangled states one needs to measure of the two states of polarization of the photons F1 and F2 plus the control measurements (B and B$_1$). The required appropriate rotations are displayed in Table \ref{T6}.

\begin{table}
\begin{center}
\begin{tabular}{|| c | c | c | c | c |}
\hline 2C & 1$E$ & Control & $AD$ & AO \\ 
\hline  $|L0\rangle_{2C}$  & $|L0\rangle_{1E}$ & $|00\rangle_{BB_{1}}$ & $\alpha|00\rangle_{AD}-\beta|11\rangle_{AD}$ & $\mathbb{I}\otimes\sigma_z\sigma_x$ \\
 
\hline  $|L0\rangle_{2C}$  & $|L1\rangle_{1E}$ & $|00\rangle_{BB_{1}}$ & $\alpha|00\rangle_{AD}+\beta|11\rangle_{AD}$ & $\mathbb{I}\otimes\sigma_x$ \\

\hline  $|L1\rangle_{2C}$  & $|L0\rangle_{1E}$ & $|00\rangle_{BB_{1}}$ & $\alpha|00\rangle_{AD}+\beta|11\rangle_{AD}$ & $\mathbb{I}\otimes\sigma_x$ \\

\hline  $|L1\rangle_{2C}$  & $|L1\rangle_{1E}$ & $|00\rangle_{BB_{1}}$ & $\alpha|00\rangle_{AD}-\beta|11\rangle_{AD}$ & $\mathbb{I}\otimes\sigma_z\sigma_x$ \\
	

\hline  $|L0\rangle_{2C}$  & $|R0\rangle_{1E}$ & $|00\rangle_{BB_{1}}$ & $\alpha|01\rangle_{AD}+\beta|10\rangle_{AD}$ & $\mathbb{I}\otimes\mathbb{I}$ \\
 
\hline  $|L0\rangle_{2C}$  & $|R1\rangle_{1E}$ & $|00\rangle_{BB_{1}}$ & $\alpha|01\rangle_{AD}-\beta|10\rangle_{AD}$ & $\mathbb{I}\otimes\sigma_z$ \\

\hline  $|L1\rangle_{2C}$  & $|R0\rangle_{1E}$ & $|00\rangle_{BB_{1}}$ & $\alpha|01\rangle_{AD}-\beta|10\rangle_{AD}$ & $\mathbb{I}\otimes\sigma_z$ \\

\hline  $|L1\rangle_{2C}$  & $|R1\rangle_{1E}$ & $|00\rangle_{BB_{1}}$ & $\alpha|01\rangle_{AD}+\beta|10\rangle_{AD}$ & $\mathbb{I}\otimes\mathbb{I}$ \\


\hline  $|R0\rangle_{2C}$  & $|L0\rangle_{1E}$ & $|00\rangle_{BB_{1}}$ & $\alpha|10\rangle_{AD}+\beta|01\rangle_{AD}$ & $\sigma_x\otimes\sigma_x$ \\

\hline  $|R0\rangle_{2C}$  & $|L1\rangle_{1E}$ & $|00\rangle_{BB_{1}}$ & $\alpha|10\rangle_{AD}-\beta|01\rangle_{AD}$ & $\sigma_z\sigma_x\otimes\sigma_x$ \\

\hline  $|R1\rangle_{2C}$  & $|L0\rangle_{1E}$ & $|00\rangle_{BB_{1}}$ & $\alpha|10\rangle_{AD}-\beta|01\rangle_{AD}$ & $\sigma_z\sigma_x\otimes\sigma_x$ \\

\hline  $|R1\rangle_{2C}$  & $|L1\rangle_{1E}$ & $|00\rangle_{BB_{1}}$ & $\alpha|10\rangle_{AD}+\beta|01\rangle_{AD}$ & $\sigma_x\otimes\sigma_x$ \\


\hline  $|R0\rangle_{2C}$  & $|R0\rangle_{1E}$ & $|00\rangle_{BB_{1}}$ & $\alpha|11\rangle_{AD}-\beta|00\rangle_{AD}$ & $\sigma_z\sigma_x\otimes\mathbb{I}$ \\
 
\hline  $|R0\rangle_{2C}$  & $|R1\rangle_{1E}$ & $|00\rangle_{BB_{1}}$ & $\alpha|11\rangle_{AD}+\beta|00\rangle_{AD}$ & $\sigma_x\otimes\mathbb{I}$ \\

\hline  $|R1\rangle_{2C}$  & $|R0\rangle_{1E}$ & $|00\rangle_{BB_{1}}$ & $\alpha|11\rangle_{AD}+\beta|00\rangle_{AD}$ & $\sigma_x\otimes\mathbb{I}$ \\

\hline  $|R1\rangle_{2C}$  & $|R1\rangle_{1E}$ & $|00\rangle_{BB_{1}}$ & $\alpha|11\rangle_{AD}-\beta|00\rangle_{AD}$ & $\sigma_z\sigma_x\otimes\mathbb{I}$ \\


\hline  $|L0\rangle_{2C}$  & $|L0\rangle_{1E}$ & $|01\rangle_{BB_{1}}$ & $\alpha|10\rangle_{AD}+\beta|01\rangle_{AD}$ & $\sigma_x\otimes\sigma_x$ \\
 
\hline  $|L0\rangle_{2C}$  & $|L1\rangle_{1E}$ & $|01\rangle_{BB_{1}}$ & $\alpha|10\rangle_{AD}-\beta|01\rangle_{AD}$ & $\sigma_z\sigma_x\otimes\sigma_x$ \\

\hline  $|L1\rangle_{2C}$  & $|L0\rangle_{1E}$ & $|01\rangle_{BB_{1}}$ & $\alpha|10\rangle_{AD}-\beta|01\rangle_{AD}$ & $\sigma_z\sigma_x\otimes\sigma_x$ \\

\hline  $|L1\rangle_{2C}$  & $|L1\rangle_{1E}$ & $|01\rangle_{BB_{1}}$ & $\alpha|10\rangle_{AD}+\beta|01\rangle_{AD}$ & $\sigma_x\otimes\sigma_x$ \\
	

\hline  $|L0\rangle_{2C}$  & $|R0\rangle_{1E}$ & $|01\rangle_{BB_{1}}$ & $\alpha|11\rangle_{AD}-\beta|00\rangle_{AD}$ & $\sigma_z\sigma_x\otimes\mathbb{I}$ \\
 
\hline  $|L0\rangle_{2C}$  & $|R1\rangle_{1E}$ & $|01\rangle_{BB_{1}}$ & $\alpha|11\rangle_{AD}+\beta|00\rangle_{AD}$ & $\sigma_x\otimes\mathbb{I}$ \\

\hline  $|L1\rangle_{2C}$  & $|R0\rangle_{1E}$ & $|01\rangle_{BB_{1}}$ & $\alpha|11\rangle_{AD}+\beta|00\rangle_{AD}$ & $\sigma_x\otimes\mathbb{I}$ \\

\hline  $|L1\rangle_{2C}$  & $|R1\rangle_{1E}$ & $|01\rangle_{BB_{1}}$ & $\alpha|11\rangle_{AD}-\beta|00\rangle_{AD}$ & $\sigma_z\sigma_x\otimes\mathbb{I}$ \\


\hline  $|R0\rangle_{2C}$  & $|L0\rangle_{1E}$ & $|01\rangle_{BB_{1}}$ & $\alpha|00\rangle_{AD}-\beta|11\rangle_{AD}$ & $\mathbb{I}\otimes\sigma_z\sigma_x$ \\

\hline  $|R0\rangle_{2C}$  & $|L1\rangle_{1E}$ & $|01\rangle_{BB_{1}}$ & $\alpha|00\rangle_{AD}+\beta|11\rangle_{AD}$ & $\mathbb{I}\otimes\sigma_x$ \\

\hline  $|R1\rangle_{2C}$  & $|L0\rangle_{1E}$ & $|01\rangle_{BB_{1}}$ & $\alpha|00\rangle_{AD}+\beta|11\rangle_{AD}$ & $\mathbb{I}\otimes\sigma_x$ \\

\hline  $|R1\rangle_{2C}$  & $|L1\rangle_{1E}$ & $|01\rangle_{BB_{1}}$ & $\alpha|00\rangle_{AD}-\beta|11\rangle_{AD}$ & $\mathbb{I}\otimes\sigma_z\sigma_x$ \\


\hline  $|R0\rangle_{2C}$  & $|R0\rangle_{1E}$ & $|01\rangle_{BB_{1}}$ & $\alpha|01\rangle_{AD}+\beta|10\rangle_{AD}$ & $\mathbb{I}\otimes\mathbb{I}$ \\
 
\hline  $|R0\rangle_{2C}$  & $|R1\rangle_{1E}$ & $|01\rangle_{BB_{1}}$ & $\alpha|01\rangle_{AD}-\beta|10\rangle_{AD}$ & $\sigma_z\otimes\mathbb{I}$ \\

\hline  $|R1\rangle_{2C}$  & $|R0\rangle_{1E}$ & $|01\rangle_{BB_{1}}$ & $\alpha|01\rangle_{AD}-\beta|10\rangle_{AD}$ & $\sigma_z\otimes\mathbb{I}$ \\

\hline  $|R1\rangle_{2C}$  & $|R1\rangle_{1E}$ & $|01\rangle_{BB_{1}}$ & $\alpha|01\rangle_{AD}+\beta|10\rangle_{AD}$ & $\mathbb{I}\otimes\mathbb{I}$ \\
\hline

\end{tabular}
\begin{tabular}{|| c | c | c | c | c |}
\hline 2C & 1$E$ & Control & $AD$ & AO \\
\hline  $|L0\rangle_{2C}$  & $|L0\rangle_{1E}$ & $|10\rangle_{BB_{1}}$ & $\alpha|10\rangle_{AD}+\beta|01\rangle_{AD}$ & $\sigma_x\otimes\sigma_x$ \\
 
\hline  $|L0\rangle_{2C}$  & $|L1\rangle_{1E}$ & $|10\rangle_{BB_{1}}$ & $\alpha|10\rangle_{AD}-\beta|01\rangle_{AD}$ & $\sigma_z\sigma_x\otimes\sigma_x$ \\

\hline  $|L1\rangle_{2C}$  & $|L0\rangle_{1E}$ & $|10\rangle_{BB_{1}}$ & $\alpha|10\rangle_{AD}-\beta|01\rangle_{AD}$ & $\sigma_z\sigma_x\otimes\sigma_x$ \\

\hline  $|L1\rangle_{2C}$  & $|L1\rangle_{1E}$ & $|10\rangle_{BB_{1}}$ & $\alpha|10\rangle_{AD}+\beta|01\rangle_{AD}$ & $\sigma_x\otimes\sigma_x$ \\
	

\hline  $|L0\rangle_{2C}$  & $|R0\rangle_{1E}$ & $|10\rangle_{BB_{1}}$ & $\alpha|11\rangle_{AD}-\beta|00\rangle_{AD}$ & $\sigma_z\sigma_x\otimes\mathbb{I}$ \\
 
\hline  $|L0\rangle_{2C}$  & $|R1\rangle_{1E}$ & $|10\rangle_{BB_{1}}$ & $\alpha|11\rangle_{AD}+\beta|00\rangle_{AD}$ & $\sigma_x\otimes\mathbb{I}$ \\

\hline  $|L1\rangle_{2C}$  & $|R0\rangle_{1E}$ & $|10\rangle_{BB_{1}}$ & $\alpha|11\rangle_{AD}+\beta|00\rangle_{AD}$ & $\sigma_x\otimes\mathbb{I}$ \\

\hline  $|L1\rangle_{2C}$  & $|R1\rangle_{1E}$ & $|10\rangle_{BB_{1}}$ & $\alpha|11\rangle_{AD}-\beta|00\rangle_{AD}$ & $\sigma_z\sigma_x\otimes\mathbb{I}$ \\


\hline  $|R0\rangle_{2C}$  & $|L0\rangle_{1E}$ & $|10\rangle_{BB_{1}}$ & $\alpha|00\rangle_{AD}-\beta|11\rangle_{AD}$ & $\mathbb{I}\otimes\sigma_z\sigma_x$ \\

\hline  $|R0\rangle_{2C}$  & $|L1\rangle_{1E}$ & $|10\rangle_{BB_{1}}$ & $\alpha|00\rangle_{AD}+\beta|11\rangle_{AD}$ & $\mathbb{I}\otimes\sigma_x$ \\

\hline  $|R1\rangle_{2C}$  & $|L0\rangle_{1E}$ & $|10\rangle_{BB_{1}}$ & $\alpha|00\rangle_{AD}+\beta|11\rangle_{AD}$ & $\mathbb{I}\otimes\sigma_x$ \\

\hline  $|R1\rangle_{2C}$  & $|L1\rangle_{1E}$ & $|10\rangle_{BB_{1}}$ & $\alpha|00\rangle_{AD}-\beta|11\rangle_{AD}$ & $\mathbb{I}\otimes\sigma_z\sigma_x$ \\


\hline  $|R0\rangle_{2C}$  & $|R0\rangle_{1E}$ & $|10\rangle_{BB_{1}}$ & $\alpha|01\rangle_{AD}+\beta|10\rangle_{AD}$ & $\mathbb{I}\otimes\mathbb{I}$ \\
 
\hline  $|R0\rangle_{2C}$  & $|R1\rangle_{1E}$ & $|10\rangle_{BB_{1}}$ & $\alpha|01\rangle_{AD}-\beta|10\rangle_{AD}$ & $\sigma_z\otimes\mathbb{I}$ \\

\hline  $|R1\rangle_{2C}$  & $|R0\rangle_{1E}$ & $|10\rangle_{BB_{1}}$ & $\alpha|01\rangle_{AD}-\beta|10\rangle_{AD}$ & $\sigma_z\otimes\mathbb{I}$ \\

\hline  $|R1\rangle_{2C}$  & $|R1\rangle_{1E}$ & $|10\rangle_{BB_{1}}$ & $\alpha|01\rangle_{AD}+\beta|10\rangle_{AD}$ & $\mathbb{I}\otimes\mathbb{I}$ \\


\hline  $|L0\rangle_{2C}$  & $|L0\rangle_{1E}$ & $|11\rangle_{BB_{1}}$ & $\alpha|00\rangle_{AD}-\beta|11\rangle_{AD}$ & $\mathbb{I}\otimes\sigma_z\sigma_x$ \\
 
\hline  $|L0\rangle_{2C}$  & $|L1\rangle_{1E}$ & $|11\rangle_{BB_{1}}$ & $\alpha|00\rangle_{AD}+\beta|11\rangle_{AD}$ & $\mathbb{I}\otimes\sigma_x$ \\

\hline  $|L1\rangle_{2C}$  & $|L0\rangle_{1E}$ & $|11\rangle_{BB_{1}}$ & $\alpha|00\rangle_{AD}+\beta|11\rangle_{AD}$ & $\mathbb{I}\otimes\sigma_x$ \\

\hline  $|L1\rangle_{2C}$  & $|L1\rangle_{1E}$ & $|11\rangle_{BB_{1}}$ & $\alpha|00\rangle_{AD}-\beta|11\rangle_{AD}$ & $\mathbb{I}\otimes\sigma_z\sigma_x$ \\
	

\hline  $|L0\rangle_{2C}$  & $|R0\rangle_{1E}$ & $|11\rangle_{BB_{1}}$ & $\alpha|01\rangle_{AD}+\beta|10\rangle_{AD}$ & $\mathbb{I}\otimes\mathbb{I}$ \\
 
\hline  $|L0\rangle_{2C}$  & $|R1\rangle_{1E}$ & $|11\rangle_{BB_{1}}$ & $\alpha|01\rangle_{AD}-\beta|10\rangle_{AD}$ & $\mathbb{I}\otimes\sigma_z$ \\

\hline  $|L1\rangle_{2C}$  & $|R0\rangle_{1E}$ & $|11\rangle_{BB_{1}}$ & $\alpha|01\rangle_{AD}-\beta|10\rangle_{AD}$ & $\mathbb{I}\otimes\sigma_z$ \\

\hline  $|L1\rangle_{2C}$  & $|R1\rangle_{1E}$ & $|11\rangle_{BB_{1}}$ & $\alpha|01\rangle_{AD}+\beta|10\rangle_{AD}$ & $\mathbb{I}\otimes\mathbb{I}$ \\


\hline  $|R0\rangle_{2C}$  & $|L0\rangle_{1E}$ & $|11\rangle_{BB_{1}}$ & $\alpha|10\rangle_{AD}+\beta|01\rangle_{AD}$ & $\sigma_x\otimes\sigma_x$ \\

\hline  $|R0\rangle_{2C}$  & $|L1\rangle_{1E}$ & $|11\rangle_{BB_{1}}$ & $\alpha|10\rangle_{AD}-\beta|01\rangle_{AD}$ & $\sigma_z\sigma_x\otimes\sigma_x$ \\

\hline  $|R1\rangle_{2C}$  & $|L0\rangle_{1E}$ & $|11\rangle_{BB_{1}}$ & $\alpha|10\rangle_{AD}-\beta|01\rangle_{AD}$ & $\sigma_z\sigma_x\otimes\sigma_x$ \\

\hline  $|R1\rangle_{2C}$  & $|L1\rangle_{1E}$ & $|11\rangle_{BB_{1}}$ & $\alpha|10\rangle_{AD}+\beta|01\rangle_{AD}$ & $\sigma_x\otimes\sigma_x$ \\


\hline  $|R0\rangle_{2C}$  & $|R0\rangle_{1E}$ & $|11\rangle_{BB_{1}}$ & $\alpha|11\rangle_{AD}-\beta|00\rangle_{AD}$ & $\sigma_z\sigma_x\otimes\mathbb{I}$ \\
 
\hline  $|R0\rangle_{2C}$  & $|R1\rangle_{1E}$ & $|11\rangle_{BB_{1}}$ & $\alpha|11\rangle_{AD}+\beta|00\rangle_{AD}$ & $\sigma_x\otimes\mathbb{I}$ \\

\hline  $|R1\rangle_{2C}$  & $|R0\rangle_{1E}$ & $|11\rangle_{BB_{1}}$ & $\alpha|11\rangle_{AD}+\beta|00\rangle_{AD}$ & $\sigma_x\otimes\mathbb{I}$ \\

\hline  $|R1\rangle_{2C}$  & $|R1\rangle_{1E}$ & $|11\rangle_{BB_{1}}$ & $\alpha|11\rangle_{AD}-\beta|00\rangle_{AD}$ & $\sigma_z\sigma_x\otimes\mathbb{I}$ \\
\hline 
\end{tabular}
\end{center}
\caption{Possible results and rotations for completing the controlled teleportation with two controls, corresponding to case \textit{ii}). The first column (2C) shows the possible results of measurements on the states of atom $C$ and photon $2$. Second column refers to the Alice's possible results of measurements on photon $1$ and atom $E$. Column $Control$ shows the state of controls $B$ and $B1$. $AD$ column does the same for the teleported state $AD$. Fourth column ($AO$) shows the corresponding Pauli matrices representing unitary operations on the atomic state required to
complete the teleportation process.}
\label{T6}
\end{table}

\textit{iii) Generalization}: the generalization of the controlled teleportation of entangled states can be done by using either a QWP$_{1}$ or a QWP$_{2}$ in the lower branch (see Fig. 7) when the scheme is implemented using odd or even number of controls, respectively.

\section{Conclusions}

In summary we presented three schemes to realize controlled teleportation of atomic states via photonic Faraday rotations in lossy optical cavities connected by optical fibers. The schemes only involve virtual excitations of the atoms and considers low-Q cavities, ideal photodetectors, and fibers without absorption. On the other hand, the practical experimental imperfections due to photon loss and inefficiency in the detectors turn the protocol as being probabilistic. In this respect, we can estimate the success probability of the scheme take into account the losses mentioned above, based in Ref. \cite{OlmschenkSC09}, e.g., considering the coupling and transmission of the photon through the single-mode optical fiber given by $T_f=0.2$, the transmission of each photon through the other optical components by $T_o=0.95$, the fraction of photons with the correct polarization $p_{\pi} = 0.5$, the quantum efficiency of the single-photon detector as $\eta=0.15$, $\Delta\Omega/4\pi=0.02$ as the solid angle of light collection, and a single-photon rate by source given by $75$kHz. So, we estimate the success probability of the CT of superposition and CPT as $P=p_{Bell} \times T_f\times T_o\times p_{\pi} \times \eta \times \Delta\Omega/4\pi \simeq 7.125\times 10^{-5}$ (considering $p_{Bell}=0.25$ as the probability of the ideal Bell-state measurement without necessity of additional rotations), which results in one successful controlled teleportation event every $\simeq0.19$s. The same estimative is obtained for the preparation of the two atoms entangled state. In the case of TC of entanglement we have two fibers branches, two photon sources and two photodetectors. So, we estimate the success probability as $P=p_{Bell} \times |T_f\times T_o\times p_{\pi} \times \eta \times \Delta\Omega/4\pi|^{2} \simeq 2.031\times 10^{-8}$, which results in one successful event every $\simeq11$min.

\textbf{\textrm{\subsection*{Acknowledgement}}}

We thank the CAPES, CNPq, FUNAPE-GO, and INCT-IQ, Brazilian agencies, for the partial supports.

\section*{References}


\end{document}